\title{On the Applicability of the Geodesic Deviation Equation in General Relativity}

\author{Dennis Philipp\footnote{Email: dennis.philipp@zarm.uni-bremen.de}, Dirk Puetzfeld\footnote{Email: dirk.puetzfeld@zarm.uni-bremen.de, URL: http://puetzfeld.org}, Claus L{\"a}mmerzahl\footnote{Email: dennis.philipp@zarm.uni-bremen.de}, \\[5pt] 
ZARM\\
University of Bremen\\
Am Fallturm, 28359 Bremen, Germany}

\date{\today}

\documentclass[11pt,english]{article}
\usepackage{graphicx}
\usepackage{psfrag}
\usepackage[pdf]{pstricks}
\usepackage{amsmath,amsfonts,amssymb}
\usepackage{tabularx}
\usepackage{rotating}
\usepackage{geometry}

\newcommand{\Weierstrass}{\raisebox{.15\baselineskip}{\Large\ensuremath{\wp}}}

\begin{document}
\maketitle

\begin{abstract}
Within the theory of General Relativity, we study the solution and range of applicability of the standard geodesic deviation equation in highly symmetric spacetimes. In the Schwarzschild spacetime, the solution is used to model satellite orbit constellations and their deviations around a spherically symmetric Earth model. We investigate the spatial shape and orbital elements of perturbations of circular reference curves. In particular, we reconsider the deviation equation in Newtonian gravity and then determine relativistic effects within the theory of General Relativity by comparison. The deviation of nearby satellite orbits, as constructed from exact solutions of the underlying geodesic equation, is compared to the solution of the geodesic deviation equation to assess the accuracy of the latter. Furthermore, we comment on the so-called Shirokov effect in the Schwarzschild spacetime and limitations of the first order deviation approach.
\end{abstract}


\section{Introduction \label{sec_introduction}}

Applications in space based geodesy and gravimetry missions require the precise knowledge of satellite orbits and possible deviations of nearby ones. In such a context, one of the satellites may serve as the reference object and measurements are performed w.r.t.\ this master spacecraft. GRACE-FO, the successor of the long-lasting GRACE mission, aims at measuring the change of the separation between two spacecrafts with some 10\ nm accuracy \cite{Flechtner:2016, Loomis:2012}. The change of this distance is then used to obtain information about the gravitational field of the Earth, i.e.\ to measure the Newtonian multipole moments of the gravitational potential, and to deduce information about the mass distribution and its temporal variations. 

In this work, we investigate the geodesic deviation equation in General Relativity (GR) in the case of highly symmetric spacetimes. Our aim is to develop a measure for the quality of approximation that the deviation equation provides to model test bodies and their orbit deviations in different orbital configurations. To achieve this, we construct general solutions of the geodesic deviation equation and compare them to exact solutions of the underlying geodesic equation in a Schwarzschild spacetime. Our analysis allows us to reveal physical and artificial effects of such an approximative description. In particular, we comment on an effect that was reported for the first time in 1973 by Shirokov \cite{Shirokov:1973}.

The structure of the paper is as follows: In section \ref{sec_Newton}, we reconsider the deviation equation in Newtonian gravity. This is followed by an investigation of the first order geodesic deviation equation in static, spherically symmetric spacetimes in GR in section \ref{sec_GR}. A direct comparison between the solution for the Schwarzschild spacetime, which we focus on for the rest of this work, and the Newtonian results unveils relativistic effects. 
In section \ref{sec_applicability}, we describe the shape of perturbed orbits and the influence of six free integration parameters on the general solution. These parameters are connected to orbital elements of the orbit under consideration, and they determine how it is obtained from a perturbation of the reference curve. We assess the range of applicability of the deviation equation by comparing its solutions to deviations constructed ``by hand'' from exact solutions of the underlying geodesic equation. We study physical effects such as perigee precession and the redshift due to time dilation between the reference and perturbed orbit.
Building on these results, we uncover some artificial effects previously reported in the literature. Our conclusions and a brief outlook on future applications are given in section \ref{sec_conclusion}. Appendix \ref{app_conventions} contains a summary of our notations and conventions.


\section{Orbit Deviations in Newtonian Gravity \label{sec_Newton}}

There are two basic equations that govern Newtonian gravitational physics; the field equation, also known as Poisson's equation
\begin{subequations}
	\begin{align}
		\Delta U(x) = \partial_\mu\partial^\mu U(x) = 4 \pi G \rho(x) \, ,
		\label{eq_NewtonFieldEq}
	\end{align}
and the equation of motion 
	\begin{align}
		\ddot{x}^\mu = - \partial^\mu U(x) \, ,
		\label{eq_NewtonEOM}
	\end{align}
\end{subequations}
where the $(x^\mu) = (x,y,z)$ are Cartesian coordinates and the overdot denotes derivatives w.r.t.\ the Newtonian absolute time $t$. Here and in the following, Greek indices are spatial indices and take values $1,2,3$. The field equation \eqref{eq_NewtonFieldEq} relates the Newtonian gravitational potential $U$ to the mass density $\rho$ and introduces Newton's gravitational constant $G$ as a factor of proportionality. Outside a spherically symmetric (and static) mass distribution, i.e.\ in the region where $\rho = 0$, we obtain as a solution of the Laplace equation $\Delta U = 0$:
\begin{align}
	U(r) = -GM/r \, , 
	\label{eq_NewtonPotential}
\end{align}
where $M$ is the mass of the central object, obtained by integrating the mass density over the three-volume of the source, and $r$ is the distance to the center of the gravitating mass. The equation of motion \eqref{eq_NewtonEOM} describes how point particles move in the gravitational potential given by $U$.


\subsection{Newtonian Deviation Equation}

We now recall the derivation of the Newtonian deviation equation, see, e.g., Ref.\  \cite{Philipp:etal:2015} and references therein. For a given reference curve $Y^\mu (t)$ that fulfills the equation of motion we construct a second curve $X^\mu(t) = Y^\mu(t) + \eta^\mu(t)$ and introduce the deviation $\eta$. This second curve shall be a solution of the equation of motion as well (at least up to linear order, as we will see below). Hence, we get
\begin{align}
	\ddot{X}^\mu = \ddot{Y}^\mu + \ddot{\eta}^\mu = - \partial^\mu U(X) = - \partial^\mu U(Y + \eta) \, .
\end{align}
For small deviations we linearize the potential around the reference object with respect to the deviation,
\begin{align}
	U(X) = U(Y+\eta) = U(Y) + \eta^\nu \partial_\nu U(Y) + \mathcal{O}(\eta^2) \, .
\end{align}
Thereupon, the first order deviation equation in Newtonian gravity becomes (spatial indices are raised and lowered with the Kronecker delta $\delta^{\mu}_{\nu}$)
\begin{align}
	\ddot{\eta}^\mu = - [\partial^\mu \partial_\nu U(Y)] \, \eta^\nu =: K^\mu{}_{\nu} \eta^\nu \, . 
	\label{eq_NewtonDevEq}
\end{align}
For a homogeneous $(\partial_\nu U \equiv 0)$ or vanishing $(U \equiv 0)$ gravitational potential, the deviation vector has the simple linear time dependence $\eta^\mu (t) = A^\mu t + B^\mu$, with constants $A^\mu$ and $B^\mu$. A non-linear time dependence of the deviation is caused by second derivatives of the Newtonian gravitational potential, i.e.\ if $K^\mu{}_{\nu} \neq 0$.

Since we are interested in the deviation for highly symmetric situations, we now use the potential \eqref{eq_NewtonPotential} outside a spherically symmetric mass distribution and introduce  usual spherical coordinates by 
\begin{align}
(x,y,z) = (r\sin\vartheta \cos\varphi, r\sin\vartheta \sin\varphi, r \cos\vartheta) \, .
\end{align}
Due to the symmetry of the situation we can, without loss of generality, restrict the reference curve to lie within the equatorial plane that is defined by $\vartheta = \pi/2$. Applying the coordinate transformation $x^a \to \tilde{x}^a$ from Cartesian to spherical coordinates, Eq.\ \eqref{eq_NewtonDevEq} turns into
\begin{align}
	\ddot{\tilde{\eta}}^\nu \tilde{\partial}_\nu x^\mu 
	+ 2 \dot{\tilde{\eta}}^\nu \tilde{\partial}_\nu \dot{x}^\mu 
	+ \tilde{\eta}^\nu \tilde{\partial}_\nu \ddot{x}^\mu
	= \left[ \partial_r U(r) ~ \partial^\mu \partial_\nu ~ r 
	+ \partial_r^2 U(r) (\partial^\mu r) (\partial_\nu r) \right] \tilde{\eta}^\sigma \tilde{\partial}_\sigma x^\nu \, ,
\end{align}
where $(\tilde{\eta}^\mu) = (\eta^r,\eta^\vartheta,\eta^\varphi)$ are the components of the deviation in the new coordinates and $(\tilde{\partial}_\mu) = (\partial_r,\partial_\vartheta,\partial_\varphi)$. These are three equations for the three unknown components of the deviation. All angular terms can be eliminated by appropriate combinations of these equations and a straightforward but rather lengthy calculation yields the system of differential equations
\begin{subequations}
	\label{eq_NewtonDevSystem}
	\begin{align}
		\ddot{\eta}^\vartheta 	&= - \frac{\dot{R}}{R} \dot{\eta}^\vartheta - \left( \frac{GM}{R^3} + \frac{\ddot{R}}{R} \right) \eta^\vartheta \, , \\
		\ddot{\eta}^r 			&= \left( \dot{\Phi}^2 + \frac{2GM}{R^3} \right) \eta^r + \left(2 \dot{R} \dot{\Phi} + R \ddot{\Phi} \right) \eta^\varphi + 2R \dot{\Phi} \, \dot{\eta}^\varphi   \, ,\\
		\ddot{\eta}^\varphi 	&= -\frac{2\dot{R}}{R} \, \dot{\eta}^\varphi - \frac{2\dot{\Phi}}{R} \dot{\eta}^r - \left( \frac{\ddot{R}}{R} + \frac{GM}{R^3} - \dot{\Phi}^2 \right)  \eta^\varphi - \frac{\ddot{\Phi}}{R} \eta^r  \, ,
	\end{align}
\end{subequations}
where the quantities represented by capital letters $(R,\Phi)$ are in general functions of time $t$ and describe the trajectory of the reference object, along which the system \eqref{eq_NewtonDevSystem} must be solved. In geodesy, the quantity $K^\mu{}_{\nu} \neq 0$ and the Eq.\ \eqref{eq_NewtonDevEq} are known in the framework of gradiometry. However, we did not find the system of differential equations \eqref{eq_NewtonDevSystem}, describing the Newtonian deviations from a general reference curve, published elsewhere in this form.


\subsection{Deviation from Circular Reference Curves}
One particular case is the deviation from a circular reference orbit with constant radius $R$. This special situation was already considered by Greenberg \cite{Greenberg:1974}, who derived (only) the oscillating solutions. However, in \cite{Philipp:etal:2015} the full solution for this case can be found. In the following we briefly summarize the results in a form that we will use later to compare to the relativistic results. The azimuthal motion of the reference orbit is described by
\begin{align}
	\dot{\Phi} = \sqrt{\dfrac{GM}{R^3}} ~ \Rightarrow ~ \Phi(t) = \sqrt{\dfrac{GM}{R^3}} \, t =: \Omega_K \, t \, .
\end{align}
The quantity $\Omega_K$ is the well known Keplerian frequency and leads to the Keplerian orbital period $2\pi / \Omega_K$. For the circular reference orbit, the conditions $\ddot{R} = \dot{R} = \ddot{\Phi}\equiv 0$ hold, and the system \eqref{eq_NewtonDevSystem} yields three ordinary second order differential equations of which the last two are coupled (see Eqns.\ (29), (33) and (34) in \cite{Greenberg:1974} for comparison)
\begin{subequations}
	\begin{align}
		\ddot{\eta}^\vartheta 	&= - \Omega_K^2 \, \eta^\vartheta \, , \\
		\ddot{\eta}^r 			&= 2 R \, \Omega_K \,\dot{\eta}^\varphi + 3\Omega_K^2 \, \eta^r \, , \\
		R \, \ddot{\eta}^\varphi 	&= - 2\Omega_K \,\dot{\eta}^r \, .
	\end{align}
\end{subequations}
The first equation for the deviation in the $\vartheta$-direction describes a simple harmonic oscillation around the reference orbital plane and is decoupled from the remaining ones. The general real-valued solution is given by
\begin{subequations}
\label{eq_NewtonDevSol}
\begin{align} 
		\eta^\vartheta(t) = \dfrac{C_{(5)}}{R} \cos \Omega_K t + \dfrac{C_{(6)}}{R} \sin \Omega_K t \, .
\end{align}
The parameters $C_{(5)}$ and $C_{(6)}$ are the amplitudes of the two fundamental solutions (normalized to the reference radius $R$) and the deviation component $\eta^\vartheta$ oscillates with the Keplerian frequency $\Omega_K$. In \cite{Greenberg:1974}, Greenberg derived the oscillating solutions for the remaining two equations. However, the general solution, cf.\ \cite{Philipp:etal:2015}, is given by 
\begin{align}
	\eta^r (t)		&= C_{(1)} + C_{(2)} \sin \Omega_K t + C_{(3)} \cos \Omega_K t \, , \\
	R \, \eta^\varphi(t) &= 2 \left(C_{(2)} \cos \Omega_K t - C_{(3)} \sin \Omega_K t\right) - \frac{3}{2} \, \Omega_K C_{(1)} t + C_{(4)} \, .
\end{align}
\end{subequations}
Summarizing the results, the perturbed orbit is described by
\begin{subequations}
\label{eq_NewtonDevSolOrbit}
\begin{align}
	r(t) 			&= R + \eta^r(t) = R + C_{(1)} + C_{(2)} \sin \Omega_K t + C_{(3)} \cos \Omega_K t, \\
	\varphi(t) 		&= \Omega_K \, t + \eta^\varphi(t) = \Omega_K \left(1 - \frac{3}{2R} C_{(1)} \right) t + \dfrac{C_{(4)}}{2R} \notag \\
					&+ \dfrac{2}{R} \left(C_{(2)} \cos \Omega_K t - C_{(3)} \sin \Omega_K t\right), \\
	\vartheta(t) 	&= \frac{\pi}{2} + \eta^\vartheta(t) = \pi/2 + \dfrac{C_{(5)}}{R} \cos \Omega_K t + \dfrac{C_{(6)}}{R} \sin \Omega_K t \, .
\end{align}
\end{subequations}
Obviously, there are several possibilities to perturb the reference orbit. The parameters $C_{(i)}, i=1 \dots 6$ define the initial position and velocity (or the orbital elements) of the test body that follows the perturbed curve. The meaning of these parameters and their impact on the perturbed orbit was studied briefly in Ref. \cite{Philipp:etal:2015}, and the analysis will be extended in section \ref{sec_applicability} in the context of the general relativistic results. Note that the only frequency appearing in the solution so far is the Keplerian frequency $\Omega_K$.


\section{Geodesic Deviation in General Relativity \label{sec_GR}}

\begin{figure}[htb]
	\centering
	\includegraphics[width=0.6\textwidth]{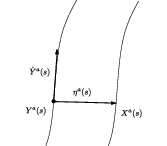}
	\caption{\label{fig_devSketch} Sketch of the deviation of two nearby geodesics $Y^a$ and ${X^a = Y^a + \eta^a}$. Here we depict the case of the orthogonal correspondence -- in which the deviation vector $\eta^a$ is chosen to be orthogonal to the four-velocity $\dot{Y}^a$ along the reference geodesic.}
\end{figure}

In GR, the equation of motion for structureless test bodies takes the form of the geodesic equation
\begin{align}
	\frac{d^2x^a}{ds^2} = - \Gamma_{bc}{}^a(x) \frac{dx^b}{ds} \frac{dx^c}{ds} \,. 
	\label{eq_geodesicEquation}
\end{align}
See Refs.\ \cite{Dixon:2015,Obukhov:Puetzfeld:2015:1} for reviews of methods to derive this, and higher order equations of motion, by means of multipolar techniques. Latin indices denote spacetime indices, taking values $0,1,2,3$, and $\Gamma_{bc}{}^a$ are the connection coefficients of the underlying spacetime (Christoffel symbols).

As in the Newtonian case, we consider two neighboring curves $Y^a(s)$ and $X^a(s)$, both of them are now assumed to be geodesics, and $s$ is the proper time measured along the curve $Y^a(s)$. Choosing $Y^a(s)$ as the reference curve, we may introduce, in a coordinate representation, the deviation $\eta^a(s)$ w.r.t.\ the neighboring curve $X^a(s)$ as
\begin{align}
	\eta^a (s) := X^a (s) - Y^a (s)  \label{eq_GRdev} \, .
\end{align}
Denoting the normalized four-velocity along the reference curve by $\dot{Y}^a:=\mathrm{d}Y^a/\mathrm{d}s$, it can be shown that the second covariant derivative of the deviation fulfills 
\begin{align}
	\dfrac{\mathrm{D}^2 \eta^a (s)}{\mathrm{d} s^2} = R^{a}{}_{bcd}(Y) \, \dot{Y}^b  \eta^c \dot{Y}^d + {\cal O}(\eta^2)\, , \label{eq_JacobiEquation}
\end{align}
up to the linear order in the deviation and its first derivative, along the reference curve. This is the well-known geodesic deviation or Jacobi equation, in which $R^{a}{}_{bcd}$ denotes the curvature of spacetime. For more details on its systematic derivation, in particular its possible generalizations, and an overview of the literature see Ref.\ \cite{Puetzfeld:Obukhov:2016:1}. From Eq.\ \eqref{eq_JacobiEquation}, we infer that the deviation $\eta^a$ will have a non-linear time dependence if and only if the spacetime is curved. For vanishing curvature, the deviation can only grow linearly in time as in the Newtonian situation for $K^\mu_{~\nu} = 0$. The Newtonian quantity $K^\mu_{~\nu}$ that measures second derivatives of the gravitational potential is replaced by the curvature tensor in GR.

In the following, we will focus on the solutions of Eq.\ \eqref{eq_JacobiEquation} in the case of timelike geodesics that may correspond to satellite orbits around the Earth. In particular, we are going to assume an orthogonal parametrization, see section III in \cite{Puetzfeld:Obukhov:2016:1}, in which the deviation is orthogonal to the velocity along the reference curve $\eta_a \dot{Y}^a=0$ -- cf.\ Fig.\ \ref{fig_devSketch} for a sketch. We will further assume the reference curve $Y^a$ to be a circular geodesic and construct orbits out of its perturbation.


\subsection{Deviation Equation in Spherically Symmetric and Static Spacetimes \label{sec_JacobiEqSol}}

In a spherically symmetric and static spacetime that is described by the metric
\begin{align}
	ds^2 = A(r) dt^2 - B(r) dr^2 - r^2 (d\vartheta^2 + \sin^2 \vartheta d\varphi^2) \, , \label{eq_MetricSphSym}
\end{align}
we use spherical coordinates $(x^a) = (t, r, \vartheta, \varphi)$ and choose units such that the speed of light $c$ and Newton's gravitational constant $G$ are equal to one. The angles $\vartheta$ and $\varphi$ are the usual polar and azimuthal angles as in spherical coordinates and the radial coordinate $r$ is defined such that spheres at a radius $r$ have area $4\pi r^2$. In these coordinates, the reference geodesic shall be represented by $(Y^a) = (T,R,\Theta,\Phi)$. Due to the symmetry of the spacetime we can, without loss of generality, assume that the reference geodesic is confined to the equatorial plane. Thus, we have $\Theta \equiv \pi/2$ and $\dot{\Theta} = \ddot{\Theta} = 0$. For geodesics in the considered spacetime there exist constants of motion that correspond to the conservation of energy $E$ and angular momentum $L$, see for example \cite{Fuchs:1977}. Since the metric \eqref{eq_MetricSphSym} does neither depend on the time coordinate $t$ nor on the angle $\varphi$, i.e.\ $\partial_t$ and $\partial_\varphi$ are Killing vector fields, the constants of motion are given by 
\begin{align}
\label{eq_ConstantsOfMotion}
		E := A(r) \, \dot{t} = \text{const.} \, , \quad L := r^2 \, \dot{\varphi} = \text{const.}
\end{align}

The general solution of the first order geodesic deviation equation \eqref{eq_JacobiEquation} in the spacetime \eqref{eq_MetricSphSym} was given by Fuchs \cite{Fuchs:1983} in terms of first integrals, which remain to be solved. Unfortunately, this solution is not applicable to the simplest case of the  deviation from a circular reference geodesic. The condition $\dot{R} = 0$ causes singularities in terms $\sim 1/\dot{R}$ that appear in the equations. Shirokov \cite{Shirokov:1973} was the first to derive periodic solutions for the deviation from circular reference geodesics in Schwarzschild spacetime. In \cite{Kerner:etal:2001, Koekoek:etal:2011} the solution for Schwarzschild spacetime and circular reference geodesics was given in terms of relativistic epicycles. However, another possible way to obtain the full solution for circular reference geodesics in the more general spacetime \eqref{eq_MetricSphSym} is to refer the system of differential equations \eqref{eq_JacobiEquation} to a parallel propagated tetrad along the reference curve. The solution of the equations in this reference system is then projected on the coordinate basis \cite{Fuchs:1990}. This method is of direct relevance for relativistic geodesy, since it allows to describe the deviation as observed in the comoving local tetrad, i.e.\ by an observer with an orthonormal frame who is located at the position of the reference object. We will use the results of this method here.

The motion along the circular reference geodesic with radius $R$ in the equatorial plane can be described using the constants of motion $E$ and $L$ from Eq.\  \eqref{eq_ConstantsOfMotion}:
\begin{subequations}
	\begin{align}
		\Phi(s) 	&= \dot{\Phi} \, s = \frac{L}{R^2} \, s =: \Omega_\Phi\, s \, , \\
		T(s) 		&= \dot{T} \, s = \frac{E}{A(R)} \, s \, .
	\end{align}
\end{subequations}
After some lengthy calculations one arrives at the solution of the deviation equation using the result of Fuchs \cite{Fuchs:1990}
\begin{subequations}
\label{eq_SolutionDeviationGeneral}
	\begin{align}
		\eta^t(s) 			&= \frac{L E}{A \sqrt{L^2+R^2}} \, f(s) \, , \\
		\eta^r(s) 			&= \frac{E R}{\sqrt{AB}\sqrt{L^2+R^2}} \, g(s) \, , \\
		\eta^\vartheta(s) &= \frac{C_{(5)}}{R} \cos \Omega_\Phi s + \frac{C_{(6)}}{R} \sin \Omega_\Phi s \, ,\\
		\eta^\varphi(s) 	&= \frac{\sqrt{L^2+R^2}}{R^2} \, f(s) \, ,
	\end{align}
\end{subequations}
where the two proper time dependent functions $f(s)$ and $g(s)$ are given by
\begin{subequations}
	\begin{align}
		f(s)	&= \sqrt{\frac{\Delta}{k^2}} (C_{(2)} \cos ks - C_{(3)} \sin ks) + \dfrac{(k^2 - \Delta)}{\sqrt{\Delta}} C_{(1)}s + C_{(4)} \, , \\
		g(s)	&= C_{(1)} + C_{(2)} \sin ks + C_{(3)} \cos ks \, , \\
		k^2		 &:= \dfrac{2A''A - A'^2}{2AB(2A-A'R)} + \dfrac{3A'}{2AB\, R} \, \, , \\ 
		\Delta &:= \dfrac{2A'}{AB\,R} \, .
	\end{align}
\end{subequations}
The prime denotes derivatives w.r.t.\ the radial coordinate and the metric functions $A=A(R), ~B=B(R)$ are to be evaluated at the reference radius $R$. Furthermore, for a circular geodesic the constants of motion \eqref{eq_ConstantsOfMotion} can be expressed by
\begin{align}
	E^2 = \dfrac{2A^2}{2A-A'R} \, , \quad L^2 = \dfrac{R^3A'}{2A-A'R} \, .	
\end{align}


\subsection{Deviation Equation in Schwarzschild Spacetime}

In GR, the Schwarzschild spacetime serves as the simplest model of an isolated and spherically symmetric central object and might be used as a first order approximation of an astrophysical object like the Earth\footnote{Planets do not possess any net charge, therefore we do not consider charged solutions like, for example, the Reissner-Nordstr\o m spacetime.}. The metric functions in Eq.\ \eqref{eq_MetricSphSym} are then given by
\begin{align}
	A(r) = 1-\dfrac{2m}{r} \, , \quad B(r) = A(r)^{-1} \, .
\end{align}
The constants of motion $E$ and $L$ as well as the remaining quantities $k$ and $\Delta$ are uniquely defined by the radius $R$ of the circular reference geodesic
	\begin{align}
	\label{eq_ConstantsCircularReference}
		\Delta 	&= \frac{4m}{R^3} \, , \quad 			&&k^2 = \frac{m(R-6m)}{R^3(R-3m)} \, , \notag \\
		E^2 	&= \frac{(R-2m)^2}{R(R-3m)} \, , \quad &&L^2 = \frac{mR^2}{R-3m} \, .
	\end{align}
	
\begin{figure}[htb]
\centering
	\includegraphics[width=0.75\textwidth]{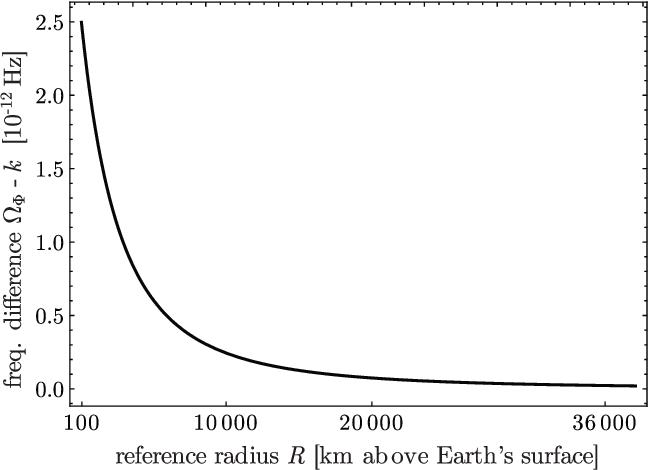}
	\caption{\label{fig_frequComparison} The difference between the frequencies $\Omega_\Phi$ and $k$, which appear in the solution of the deviation equation, is shown for reference radii that belong to satellite orbits around the Earth. The frequency difference is of the order of some $10^{-12}\,$Hz, which yields a period difference in the range of 10 - 30$\,\mu$s. We consider as the mean Earth radius $R_\oplus =6.37 \cdot 10^3$ km.}
\end{figure}

We should mention that the mass of the Earth in the units that we use is $m \approx 0.5\,$cm. The parameters $C_{(1,\dots,6)}$ will be used to model different orbital scenarios. Using the constants in Eq.\ \eqref{eq_ConstantsCircularReference}, we can simplify the solution \eqref{eq_SolutionDeviationGeneral} for the case of Schwarzschild spacetime. We find \footnote{W.r.t.\ Eq.\ \eqref{eq_SolutionDeviationGeneral} we have slightly redefined the constant parameters $C_{(1, \dots, 6)}$ in a way such that $\eta^r(s) = g(s)$. This is always possible since all coefficients preceding the functions $f(s)$ and $g(s)$ in \eqref{eq_SolutionDeviationGeneral} are constant because the reference radius $R$ is constant.}:
\begin{subequations}
\label{eq_GR_CircRefDevSol}
	\begin{align}
		\eta^t(s) 			  &= \frac{\sqrt{mR}}{R-2m} f(s) \, , \\
		\eta^r(s) 			  &= g(s) \, , \\
		\eta^\vartheta(s) &= \frac{C_{(5)}}{R} \cos \Omega_\Phi s + \frac{C_{(6)}}{R} \sin \Omega_\Phi s \, ,\\
		\eta^\varphi(s) 	&= \dfrac{f(s)}{R} \, ,
	\end{align}
\end{subequations}
where the two functions $f(s)$ and $g(s)$ are given by
\begin{subequations}
	\label{eq_GR_CircRefDevSol2}
	\begin{align}
		f(s) 	&= 2 \sqrt{\frac{R}{R-6m}} \left( C_{(2)} \cos ks - C_{(3)} \sin ks \right) -\frac{3}{2} \Omega_\Phi \dfrac{R-2m}{R-3m} \, C_{(1)} \, s + C_{(4)}\,, \\
		g(s) 	&= C_{(1)} + C_{(2)} \sin ks + C_{(3)} \cos ks \, .
	\end{align}
\end{subequations}
Notice that the function $f(s)$ contains, besides periodic and constant parts, a term that grows linearly with the reference proper time $s$ for $C_{(1)} \neq 0$. This contribution is not bounded and will, thus, limit the validity of the framework since we work with the first order deviation equation, i.e.\ the deviation $\eta^a$ is assumed to be small and only contributions up to first order were considered. We observe that in the general relativistic solution of the first order deviation equation two distinct frequencies appear:
\begin{subequations}
\label{eq_TwoFrequencies}
\begin{align}
	k 				&= \sqrt{\dfrac{m}{R^3}} \sqrt{\frac{R-6m}{R-3m}} = \Omega_K \sqrt{\frac{R-6m}{R-3m}} \, , \\
	\Omega_\Phi 	&= \sqrt{\dfrac{m}{R^3}} \sqrt{\dfrac{R}{R-3m}} = \Omega_K \sqrt{\dfrac{R}{R-3m}} \, .
\end{align}
\end{subequations}
In the Newtonian limit these two frequencies coincide and yield the Keplerian frequency $\Omega_K$. Fig.\ \ref{fig_frequComparison} shows the difference $\Omega_\Phi - k$ between both frequencies for reference radii that correspond to satellite orbits from $100$ km to $3.6\cdot 10^4$ km above the surface of the Earth. It is worthwhile to note that the general relativistic solution \eqref{eq_GR_CircRefDevSol}, \eqref{eq_GR_CircRefDevSol2} approaches the correct Newtonian limit \eqref{eq_NewtonDevSol} for $c \to \infty$. Studying the difference allows to uncover relativistic effects in the following. Observe that the normalization of the reference four-velocity yields
\begin{align}
	(1-2m/R) \, \dot{T}^2 - R^2 \dot{\Phi}^2 &= (1-2m/R) \dot{T}^2 - r^2 \Omega^2_\Phi = 1 \\ 
	&\Rightarrow \quad \dot{T} = \sqrt{\dfrac{R}{R-3m}} \, .
\end{align}
When we parametrize the circular reference orbit by coordinate time we get
\begin{align}
	\tilde{\Omega}_\Phi := \frac{d\Phi}{ds} \left(\frac{dT}{ds} \right)^{-1} = \Omega_\Phi \dot{T}^{-1} = \sqrt{\dfrac{m}{R^3}} = \Omega_K \, .
\end{align}
Hence, Kepler's third law holds perfectly well for circular orbits in the Schwarzschild spacetime when the orbit is parametrized by coordinate time.


\section{Applicability of the Geodesic Deviation Equation \label{sec_applicability}}

In this section, we study the applicability of the first order deviation equation \eqref{eq_JacobiEquation} in Schwarzschild spacetime to describe the motion of a test body that is close to a given circular reference geodesic. Its worldline is determined by a small initial perturbation of that reference curve, described by the solution \eqref{eq_GR_CircRefDevSol}. In the following, we investigate the shape of the perturbed orbits as well as physical and artificial effects, which are present in the solution. 

To describe different orbital configurations we have to examine the impact of the parameters $C_{(1,\dots,6)}$ on the perturbed orbit. A proper way to do this is to investigate the impact of each parameter separately since the different effects can be superimposed in this linearized framework. Here, we extend the brief analysis that was done in Ref. \cite{Philipp:etal:2015}. The connection between the parameters $C_{(i)}$ and the orbital elements of the perturbed orbit are summarized in Tab.\ \ref{tab_initialState}. Hence, for a specific orbital configuration that is to be modeled we can determine the parameters that must be taken into account from the table and describe that satellite configuration within the framework of the geodesic deviation equation using the solution \eqref{eq_GR_CircRefDevSol}.


\subsection{Shape of the Perturbed Orbits \label{sec_orbitShape}}
\begin{figure*}[htb]
\centering
	\includegraphics[width=0.32\textwidth]{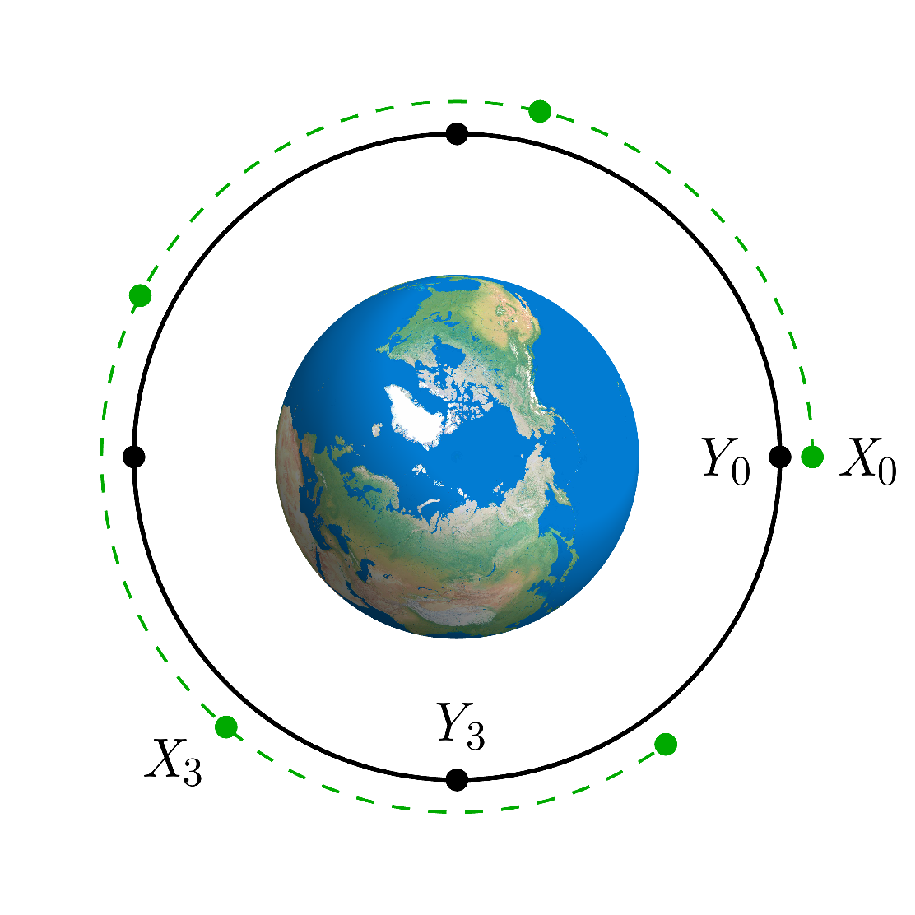} \hfill
	\includegraphics[width=0.32\textwidth]{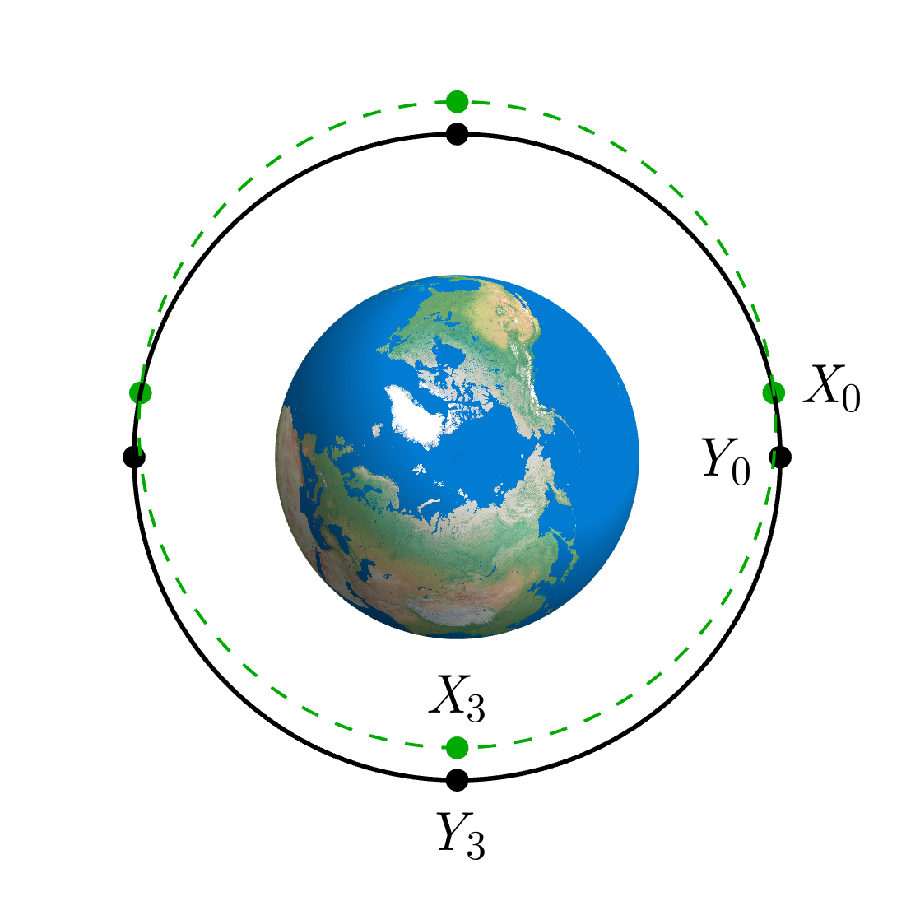} \hfill
	\includegraphics[width=0.32\textwidth]{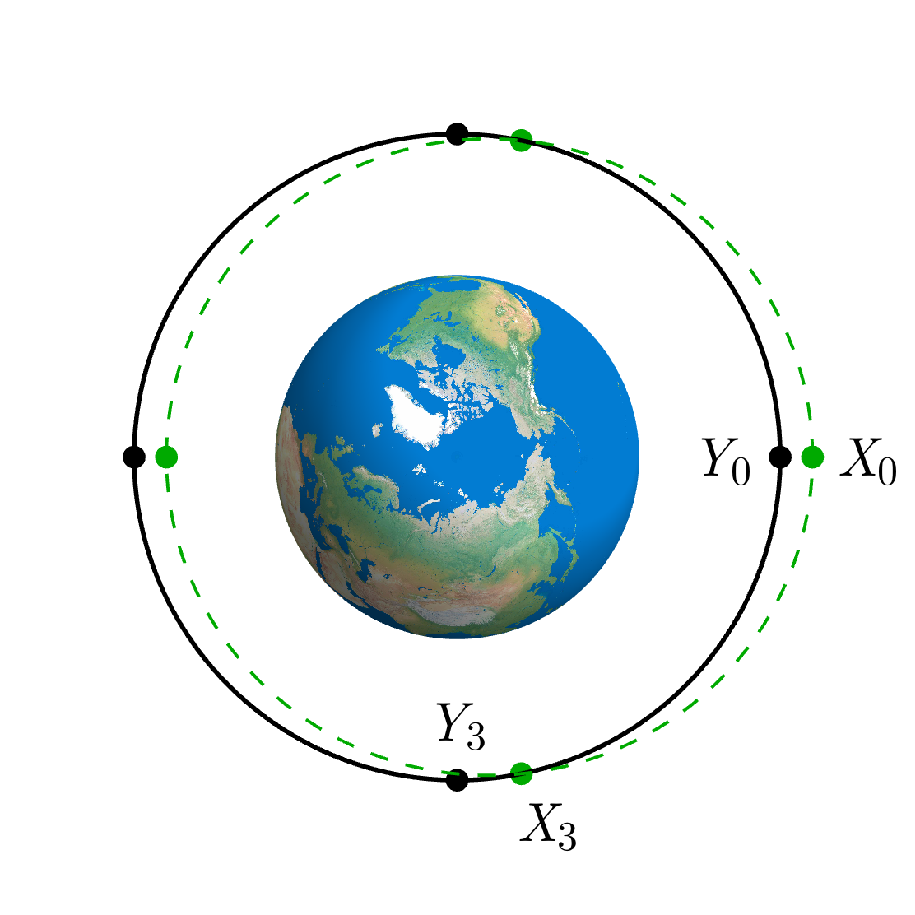} \\
	\includegraphics[width=0.32\textwidth]{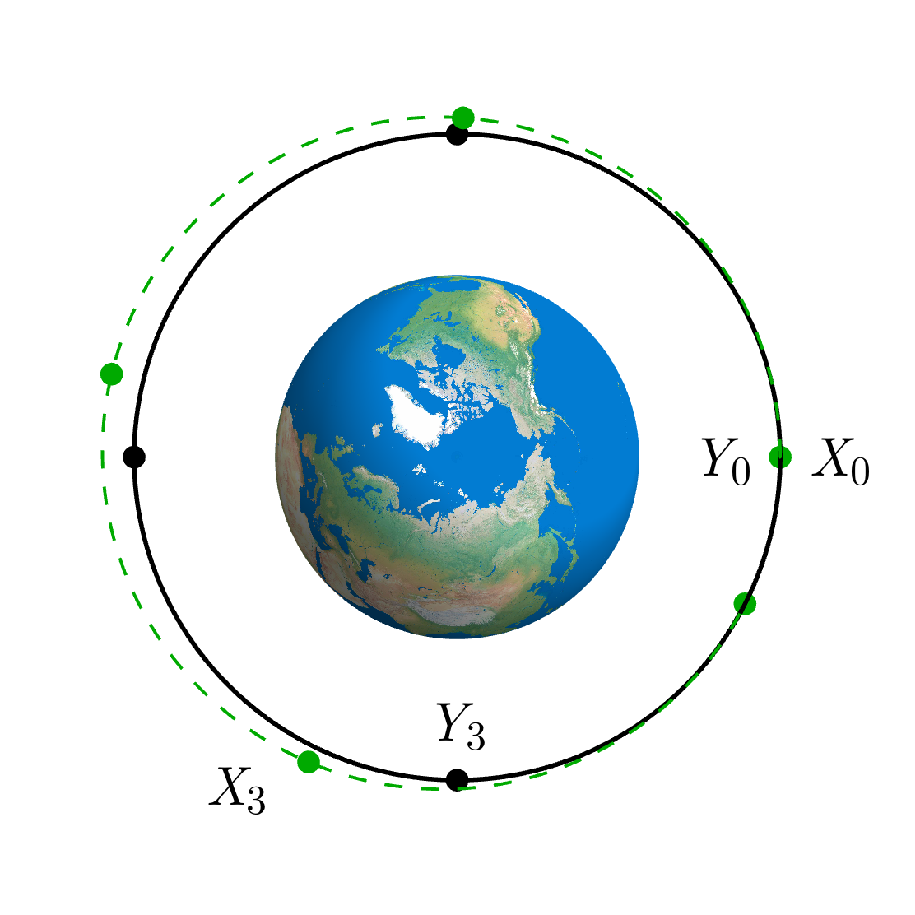} \hfill
	\includegraphics[width=0.32\textwidth]{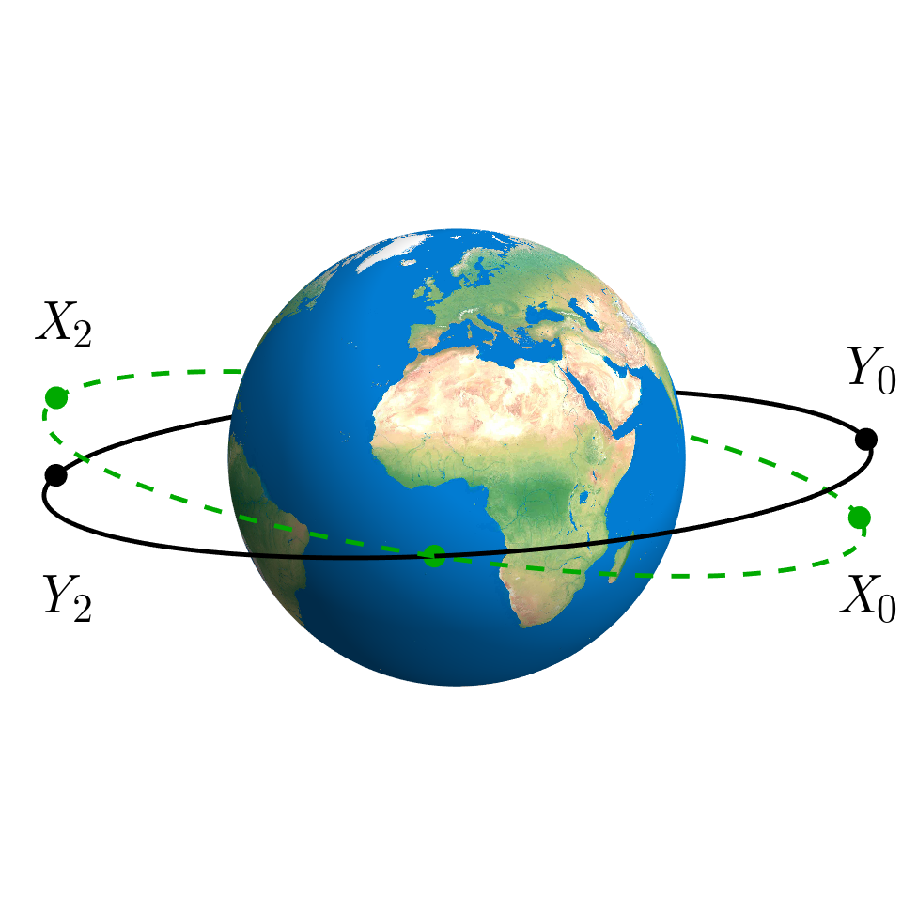} \hfill
	\includegraphics[width=0.32\textwidth]{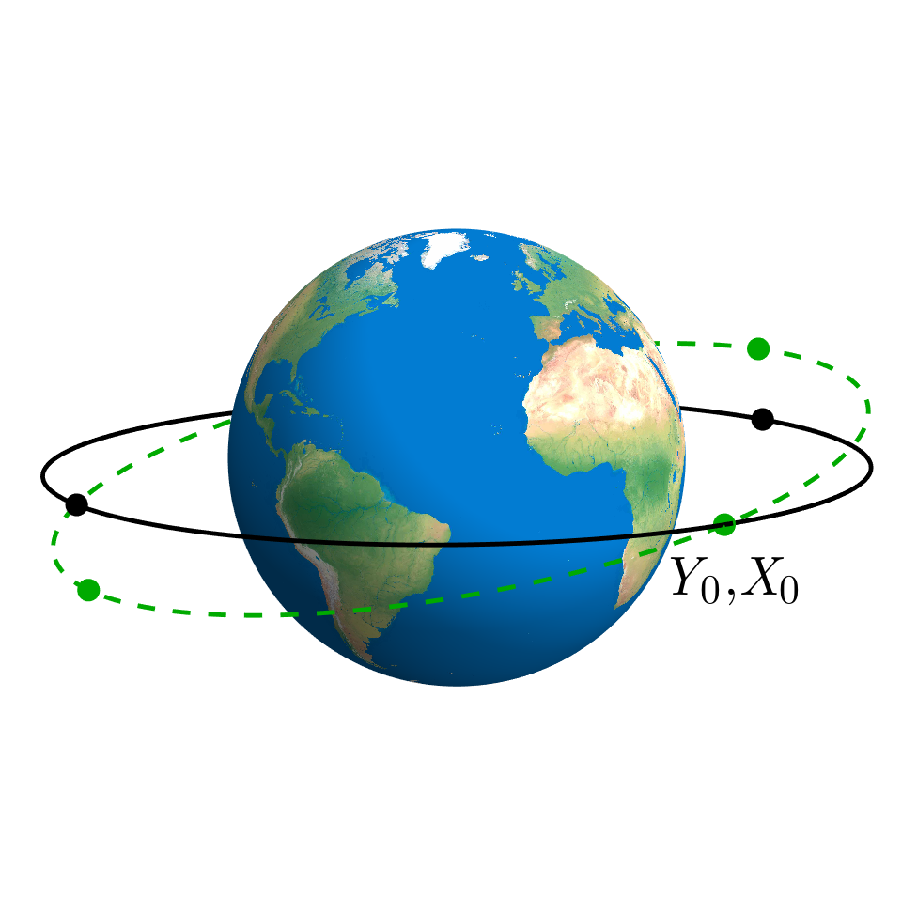} 
	\caption{\label{fig_orbitShape} The solid line (black) shows the circular reference orbit, whereas the dashed line (green) shows the perturbed orbit as calculated with the solution of the deviation equation for: only $C_{(1)} \neq 0$ (top, left), only $C_{(2)} \neq 0$ (top, middle), only $C_{(3)} \neq 0$ (top, right), a combination of both such that initially $r(0) = R, \, \dot{r}(0) = 0$ (bottom left) and a pendulum orbit as the result of an inclined perturbation using only $C_{(5)}$ (bottom middle) and $C_{(6)}$ (bottom right). We have marked the respective positions $Y_n$ on the reference orbit and $X_n$ on the perturbed orbit for reference proper time values $s = n/4 \cdot 2\pi/\Omega_\Phi$. We used a reference radius of $R = R_\oplus + 5000\,$km and a mean Earth radius $R_\oplus =6.37 \cdot 10^3$ km.}
\end{figure*}

The following sections are named after the geometric shape of the perturbed orbits, caused by the choice of the respectively considered parameter(s). All orbits that we discuss in the following are shown in Fig.\ \ref{fig_orbitShape}.


\subsubsection{Circular Perturbation}
\begin{figure}[htb]
\centering
	\includegraphics[width=0.75\textwidth]{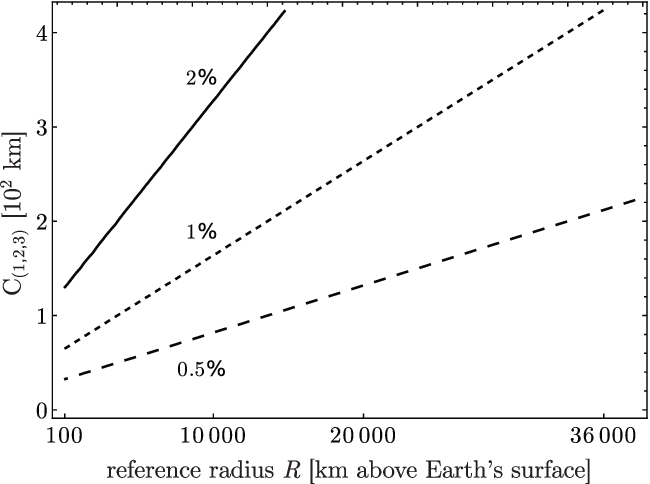} 
	\caption{\label{fig_devmagnitude} The magnitude of the normalized radial perturbation $C_{(i)} / R$ as a function of $C_{(1,2,3)}$ and the reference radius $R$. The lines represent surfaces of constant $C_{(i)} / R$. We use a mean Earth radius $R_\oplus =6.37 \cdot 10^3$ km.}
\end{figure}
If we set all parameters but $C_{(1)}$ equal to zero, the perturbed orbit remains in the reference orbital plane and has still a circular shape. This perturbed orbit is given by 
\begin{subequations}
\label{eq_circPerturb}
\begin{align}
	r			&= R + \eta^r = R + C_{(1)} \, , \\
	t(s) 		&= \dot{T} s + \eta^t(s) = \dot{T} s - \dfrac{3}{2} \dot{T} \dfrac{m}{R-3m} \, \dfrac{C_{(1)}}{R}\, s =: \left( \dot{T} + \delta t_{(1)} \right) s, \\
	\varphi(s) 	&= \Omega_\Phi \, s + \eta^\varphi(s) = \Omega_\Phi \, s - \dfrac{3}{2} \Omega_\Phi \dfrac{R-2m}{R-3m} \dfrac{C_{(1)}}{R} \, s =: (\Omega_\Phi + \delta \omega_{(1)})\, s.
\end{align}
\end{subequations}
The reference and the perturbed orbit are shown in Fig.\ \ref{fig_orbitShape} for one reference period and a chosen reference radius of $5000\,$km above the surface of the Earth. As one would expect, a positive radial perturbation $C_{(1)}$ yields a smaller azimuthal frequency, $ \dot{\varphi} = \Omega_\Phi + \delta \omega_{(1)} < \Omega_\Phi$, as compared to the reference motion. The frequency and radial perturbations are related via
\begin{align}
\label{eq_circPerturbRelation}
	\dfrac{\delta \omega_{(1)}}{\Omega_\Phi} = -\dfrac{3}{2} \dfrac{R-2m}{R-3m} \dfrac{C_{(1)}}{R} \, ,
\end{align}
such that they are not independent. Since we work with the first order deviation equation, we have to ensure that the radial perturbation is indeed small, i.e.\ $C_{(1)}/R \ll 1$. This is related to upper bounds for $C_{(1)}$ that need to be chosen in a proper way. For a satellite orbit of about $10^4$ km above the surface of the Earth, the normalized perturbation is $C_{(1)}/R \approx 10^{-7}\text{m}^{-1} \, C_{(1)}$. Hence, the allowed values for $C_{(1)}$ strongly depend on the chosen reference radius and given upper bounds for the radial perturbation. For various values of the reference radius - ranging from Low Earth Orbits (LEO) to geostationary ones - and the parameter $C_{(1)}$, we show the magnitude of the normalized radial perturbation $C_{(1)} / R$ in Fig.\ \ref{fig_devmagnitude}. To decide whether the description of a satellite configuration within the framework of the first order deviation equation is useful or not, one has to define the reference radius and the maximal radial perturbation for the desired scenario. The value $C_{(1)}/R$ can then be estimated from Fig.\ \ref{fig_devmagnitude} and if it fulfills the condition $C_{(1)}/R \ll 1$ the solution may give a simple and useful description.

After a full azimuthal period on the reference orbit, $s = 2\pi / \Omega_\Phi$, and the perturbed orbit is not yet closed since $\varphi(2\pi/\Omega_\Phi) \neq 2\pi$. The deficit angle $\Delta \alpha$ is given by 
\begin{align}
\label{eq_circPerturbDeficitAngle}
	\Delta \alpha &= \varphi \left( 2\pi/\Omega_\Phi \right) - 2\pi = 2\pi \dfrac{\delta \omega_{(1)}}{\Omega_\Phi} = - 3\pi \dfrac{R-2m}{R-3m} \dfrac{C_{(1)}}{R} \, .
\end{align}
This angle may correspond, in principle, to an observable quantity and the relation can be solved for $m$ explicitly,
\begin{align}
	m = \dfrac{R (3C_{(1)}\pi + \Delta \alpha R)}{3(2C_{(1)}\pi + \Delta \alpha R)} \, ,
\end{align}
to obtain an estimate for the relativistic mass monopole of the central object. Thus, the mass can be obtained from the measurement of the deficit angle, assuming the situation can be prepared with initially known reference radius $R$ and radial distance $C_{(1)}$ between both orbits. Also the deviation of a test object from the center of mass within a hollow satellite might be used for such a measurement.


\subsubsection{Elliptical Perturbation I}
The two parameters $C_{(2,3)}$ cause elliptical perturbations in the (reference orbital plane) if we neglect the influence of all other parameters, i.e.\ if $C_{(1,4,5,6)} =  0$. If we choose to have $C_{(2)} = 0$, the perturbed orbit is described by
\begin{subequations}
\label{eq_elliptPerturb}
\begin{align}
		r(s) 	&= R + \eta^r(s) = R + C_{(3)} \cos ks = R + C_{(3)} + \mathcal{O}\left(s^2\right) \, , \\
		t(s) 	&= \dot{T} s + \eta^t(s) \notag \\
				&= \dot{T} s - \dfrac{2R}{R-2m} \sqrt{\dfrac{m}{R-6m}} C_{(3)} \sin ks \notag \\
				&=: \dot{T} s + \delta t_{(3)} \sin ks = (\dot{T} + \delta t_{(3)} k)\, s + \mathcal{O}\left(s^2\right) \, , \\
	\varphi(s) 	&= \Omega_\Phi s - 2 \sqrt{\dfrac{R}{R-6m}} \dfrac{C_{(3)}}{R} \sin ks \notag \\
				&:= \Omega_\Phi s + \delta \omega_{(3)} \sin ks = (\Omega_\Phi + \delta \omega_{(3)}k)\, s + \mathcal{O}\left(s^2\right).
\end{align}
\end{subequations}
For an elliptically perturbed orbit of this kind, the eccentricity $e$ and the semi major axis $a$ can be linked to the radial perturbation via
\begin{subequations}
\begin{align}
	e = \dfrac{C_{(3)}}{R} \, , \quad a = R \, .
\end{align}
We can as well calculate the distance to the perigee $d_p$ and apogee $d_a$ that are related to the radial perturbation
\begin{align}
	d_p = R-C_{(3)} \, , \quad d_a = R+C_{(3)} \, ,
\end{align}
and confirm that the semi major axis is half of the sum of the two distances as it should be.
\end{subequations}
Using these relations the spatial shape of the perturbed orbit can be represented in a familiar way as
\begin{subequations}
\begin{align}
	r(s) 		&= a(1+e \cos ks) \, ,\\
	\varphi(s) 	&= \Omega_\Phi s - 2 \sqrt{\dfrac{a}{a-6m}} e \sin ks \, .
\end{align}
\end{subequations}
The eccentricity of the perturbed orbit $e = C_{(3)} / R$ is shown in Fig.\ \ref{fig_devmagnitude}. If for a specific satellite mission the maximal allowed eccentricity is given, we can read off upper bounds for the parameter $C_{(3)}$ from Fig.\ \ref{fig_devmagnitude}, or, vice versa: we can model an orbit with a given (small) eccentricity by choosing the necessary value for $C_{(3)}$. 

The difference between the effects of the two parameters $C_{(2,3)}$ is just a phase difference, i.e.\ a spatial rotation of $\pi/2$ of the perturbed orbit within the reference orbital plane. For the case that only $C_{(2)} \neq 0$, the perturbed orbit is described by
\begin{subequations}
\begin{align}
	r(s)		&= R + \eta^r(s) = R + C_{(2)} \sin ks = R + C_{(2)} \, k s + \mathcal{O} \left(s^2\right) \, ,\\
	t(s) 		&= \dot{T} s + \eta^t(s) \notag \\
				&= \dot{T} s + \dfrac{2R}{R-2m} \sqrt{\dfrac{m}{R-6m}} C_{(2)} \cos ks \notag \\
				&=: \dot{T} s + \delta t_{(2)} \cos ks = \dot{T}s + \delta t_{(2)} + \mathcal{O} \left(s^2\right) \, , \\
	\varphi(s) 	&= \Omega_\Phi s + 2 \sqrt{\dfrac{R}{R-6m}} \dfrac{C_{(2)}}{R} \cos ks \notag \\
				&=: \Omega_\Phi s + \delta \omega_{(2)} \cos ks = \Omega_\Phi s + \delta \omega_{(2)} + \mathcal{O} \left(s^2\right) \, .
\end{align}
\end{subequations}
Both elliptical orbits are shown in Fig.\ \ref{fig_orbitShape} and the spatial rotation as the difference between the effects of $C_{(2)}$ and $C_{(3)}$ is obvious. These orbits look closed after one reference period, but they are not (recall that the frequencies $\Omega_\Phi$ and $k$ are just slightly different). After one reference period $s = 2\pi/\Omega_\Phi$ we get
\begin{subequations}
\begin{align}
	r(2\pi/\Omega_\Phi) 		&= R + C_{(3)} \cos \dfrac{2\pi k}{\Omega_\Phi} \neq r(0) \, , \\
	\varphi(2\pi/\Omega_\Phi) 	&= 2\pi + \delta \omega_{(2)} \sin \dfrac{2\pi k}{\Omega_\Phi} \neq \varphi(0) + 2\pi \, .
\end{align}
\end{subequations}
However, the difference between the two frequencies is in the range of some $10^{-12}\,$Hz and the periods differ by about $10-25\,\mu$s for reference radii in the range from LEO to geostationary orbits. The radial motion has an actual period of $s = 2\pi/k$, i.e.\ this amount of reference proper time elapses from one perigee to the next. Hence, we get
\begin{subequations}
\begin{align}
	r(2\pi/k) &= r(0) \, , \\
	\varphi(2\pi/k) &= \dfrac{2\pi \Omega_\Phi}{k} \neq\varphi(0) + 2\pi \, .
\end{align}
\end{subequations}
Since the increase in the azimuthal angle differs from $2\pi$ after one radial period, the perigee of the elliptical orbit will precess. This precession is investigated in the next section in more detail.


\subsubsection{Elliptical Perturbation II}

Another kind of elliptical orbits can be constructed via the combination with a circular perturbation. For example, we use a combination of both, $C_{(1)}$ and $C_{(3)}$, to arrive at a perturbed orbit that initially fulfills $r(0) = R \, , \, \varphi(0) = \Phi(0) = 0$ and $\dot{r}(0) = 0$. Hence, the perturbed orbit is initially as close as possible to the reference orbit - cf.\ Fig.\ \ref{fig_orbitShape}. To construct it we need to choose $C_{(3)} = - C_{(1)}$. The perturbed orbit is then described by
\begin{subequations}
\label{eq_mixedOrbit}
\begin{align}
	r(s) 		&= R + C_{(1)} (1 - \cos ks) \, , \\
	t(s)		&= (\dot{T} + \delta t_{(1)})\, s + \delta t_{(3)} \sin ks \, , \\
	\varphi(s) 	&= (\Omega_\Phi + \delta \omega_{(1)})\, s + \delta \omega_{(3)} \sin ks \, , \\
	\vartheta	&= \Theta \equiv \pi/2	\, .		
\end{align}
\end{subequations}
For this orbit type we find the eccentricity and semi major axis to be
\begin{align}
\label{eq_mixedOrbitEcc}
	e = \dfrac{C_{(1)}}{R+C_{(1)}} \, , \quad a = R + C_{(1)} \quad \Rightarrow C_{(1)} = ae \, ,
\end{align}
and this allows to recast the radial motion, again, in the familiar way
\begin{align}
	r(s) 		&= a(1 - e \cos ks) \, .
\end{align}
This radial motion has a period of $2\pi/k$ and we obtain
\begin{subequations}
\label{eq_mixedOrbitPrecession}
\begin{align}
	r(2\pi/k) 		&= r(0) \, , \\
	\varphi(2\pi/k) &= \dfrac{2\pi (\Omega_\Phi + \delta \omega_{(1)})}{k} \neq \varphi(0) + 2\pi \, .
\end{align}
\end{subequations}
Hence, also this elliptical orbit will precess. The precession is studied in the next section in terms of the perigee advance.


\subsubsection{Azimuthal Perturbation}
The parameter $C_{(4)}$ causes an offset in the azimuthal motion, i.e.\ a constant phase difference between the reference and perturbed orbit. When only $C_{(4)} \neq 0$ the perturbed orbit is the same as the reference orbit, but the two test bodies are separated by a constant angle. The perturbed motion is then described by
\begin{subequations}
\begin{align}
	r 			&= R, \\
	t(s)		&= \dot{T}\, s + \eta^t(s) = \dot{T} s + \dfrac{\sqrt{mR}}{R-2m} C_{(4)}, \\
	\varphi(s) 	&= \Omega_\Phi s + \eta^\varphi(s) = \Omega_\Phi s + \dfrac{C_{(4)}}{R} \, ,
\end{align} 
\end{subequations}
where the constant azimuthal separation is determined by the value of $C_{(4)}/R$.


\subsubsection{Inclined Perturbation}
The two parameters $C_{(5)}, C_{(6)}$ incline the orbital plane with respect to the reference plane. If only $C_{(5)} \neq 0$ we obtain a circular orbit with radius $r = R$ and azimuthal motion $\varphi(s) = \Omega_\Phi s$, but with the polar motion given by
\begin{align}
\label{eq_inclinedPerturb}
	\vartheta(s) = \dfrac{\pi}{2} + \frac{C_{(5)}}{R} \cos \Omega_\Phi s \, .
\end{align}
Hence, $C_{(5)}/R$ determines the maximal inclination between the two orbital planes. If only $C_{(6)} \neq 0$ instead, there are just little changes: the $\cos(\Omega_\Phi s)$ becomes $\sin(\Omega_\Phi s)$ and the difference between the effects of these two parameters is simply related to a spatial rotation.


\subsection{The Orbital Elements}
\begin{figure}
	\centering
	\includegraphics[width=0.9\textwidth]{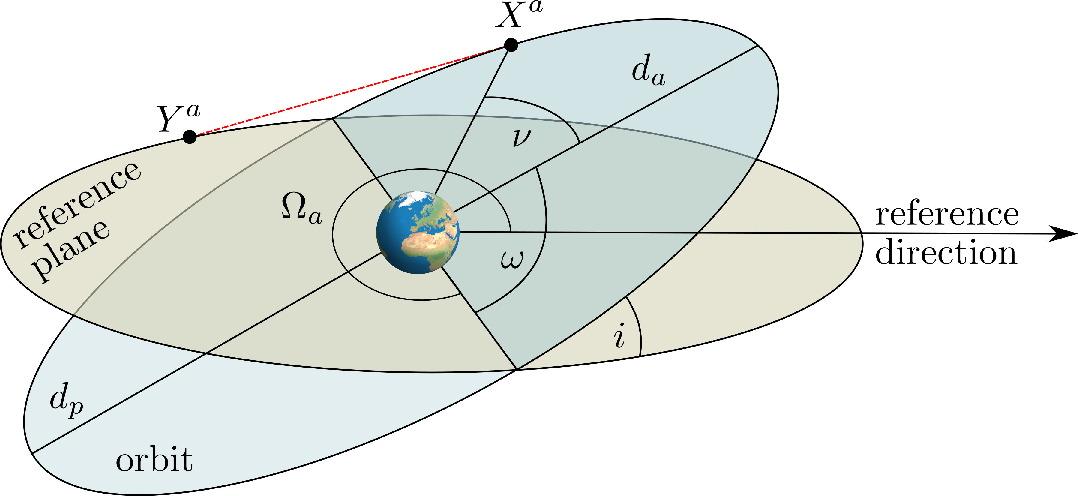}
	\caption{\label{fig_orbitalElements} A sketch of the two orbital planes including the orbital elements of the perturbed orbit $X^a$.}
\end{figure}

Combining the results of the last section we can link all parameters $C_{(1, \dots, 6)}$ to the initial position and velocity of the test body that follows the perturbed orbit. The initial quantities $r(0),\vartheta(0),\varphi(0)$ and $\dot{r}(0), \dot{\vartheta}(0), \dot{\varphi}(0)$ are summarized in Tab.\ \ref{tab_initialState} together with the resulting orbital elements of the perturbed orbit. The orbital elements are introduced in the sketch shown in Fig.\ \ref{fig_orbitalElements}.

\begin{sidewaystable}[t!]
\caption{\label{tab_initialState} The initial position and velocity of the test body that follows the perturbed orbit $X^a$. Choose one parameter that shall be the only non-zero one, then the initial state can be read off from the table. For combinations of different parameters the effects can be superimposed. We list the orbital elements such as eccentricity $e$, semi-major axis $a$, the ascending node $\Omega_{a}$ (longitude), the inclination $i$, distance to the perigee $d_p$, distance to the apogee $d_a$ and the argument of the perigee $\omega$. For the definitions of these orbital elements see Fig.\ \ref{fig_orbitalElements}. When two values are given, the orbital element depends on the sign of the respective parameter.\vspace{10pt}}
	\begin{tabular}{ccccccc}
		\hline\hline
		$\neq 0$: 				& $C_{(1)}$ 						& $C_{(2)}$ 				& $C_{(3)}$ 							& $C_{(4)}$ 		& $C_{(5)}$ 	& $C_{(6)}$ \\ \hline 	
		$r(0)$ 					& $R+C_{(1)}$ 				& $R$ 					& $R+C_{(3)}$ 					& $R$ 				& $R$ 			& $R$ \\  
		
		$\vartheta(0)$ 			& $\pi/2$ 							& $\pi/2$ 				& $\pi/2$ 								& $\pi/2$ 			& $\pi/2 + C_{(5)}/R$ & $\pi/2$ \\ 
		
		$\varphi(0)$ 			& $0$ 								& $\delta \omega_{(2)}$ & $0$ 									& $C_{(4)}/R$ 	& $0$ 			& $0$ \\ \hline 
		
		$\dot{r}(0)$ 			& $0$ 								& $C_{(2)} \, k$ & $0$ 									& $0$ 				& $0$ 			& $0$ \\ 
		
		$\dot{\vartheta}(0)$ 	& $0$ 								& $0$ 					& $0$ 									& $0$ 				& $0$ 			& $\Omega_\Phi \, C_{(6)}/R$ \\ 
		
		$\dot{\varphi}(0)$ 		& $\Omega_\Phi + \delta \omega_{(1)}$ 	& $\Omega_\Phi$ 					& $\Omega_\Phi + k ~\delta \omega_{(3)}$ 	& $\Omega_\Phi$ 			& $\Omega_\Phi$ 		& $\Omega_\Phi$ \\ \hline 
		$e$ 					& $0$ 								& $C_{(2)}/R$ 	& $C_{(3)}/R$ 					& $0$ 				& $0$ 			& $0$ \\  
		$a$ 					& $R+C_{(1)}$ 				& $R$ 					& $R$ 									& $R$ 				& $R$ 			& $R$ \\ 
		$\Omega_{a}$ 			& $0$ 								& $0$ 					& $0$ 									& $0$ 				& $\pi/2 (2- \mathrm{sgn}~C_{(5)} )$ 		& $\pi/2 (1+ \mathrm{sgn}~C_{(6)} )$ \\ \hline 
		$i$ 					& $0$ 								& $0$ 					& $0$ 									& $0$ 				& $C_{(5)}/R$ 		& $C_{(6)}/R$ \\
		$d_p$ 					& $R+C_{(1)}$ 				& $R-C_{(2)}$ 	& $R-C_{(3)}$ 					& $R$ 				& $R$ 			& $R$ \\
		$d_a$ 					& $R+C_{(1)}$ 				& $R+C_{(2)}$ 	& $R+C_{(3)}$ 					& $R$ 				& $R$ 			& $R$ \\ 
		$\omega$ 				& $0$	 							& $3\pi \Omega_\Phi/(2k) ~;~ \pi \Omega_\Phi/(2k)$ 				& $\pi \Omega_\Phi/k ~;~ 0$ 									& 	$0$ 				&  	$0$			&  $0$\\ \hline
		Shape: & circular & elliptical & elliptical & circular & circular, inclined & circular, inclined \\ \hline \hline
	\end{tabular}
\end{sidewaystable}


\subsection{Physical Effects}
\begin{figure}[htb]
	\centering
	\includegraphics[width=0.75\textwidth]{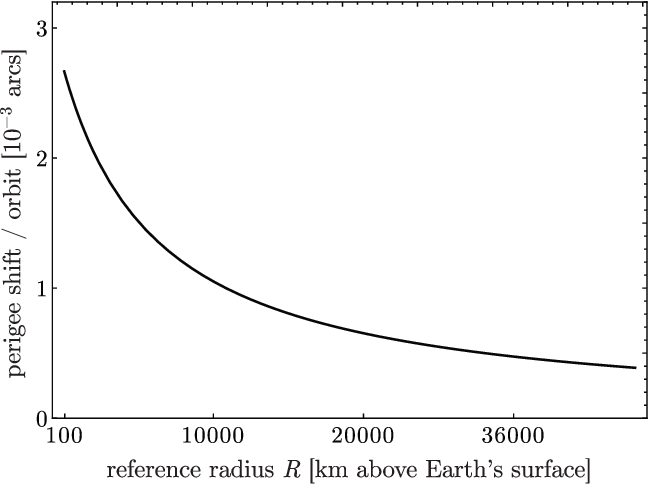}
	\caption{\label{fig_perigee} The perigee shift per orbit in $10^{-3}$ arcs for the elliptically perturbed orbit with either $C_{(3)} \neq 0$ or $C_{(2)} \neq 0$ (left) This orbit is shown in Fig.\ \ref{fig_orbitShape} in the middle of the upper row.}
\end{figure}
As shown before, the parameters $C_{(2,3)}$ lead to an elliptically perturbed orbit if at least one of them does not vanish. For such an orbit the perigee will precess and the orbit is not closed. If either of the parameters $C_{(2,3)} \neq 0$ the radial motion has a period given by $s=2 \pi/k$, but the azimuthal oscillation is advanced already. The difference to a full revolution is then given by
\begin{align}
\label{eq_perigeePre1}
	\Delta \varphi 	&= \varphi(2\pi/k) - 2\pi 	= 2\pi \left( \dfrac{\Omega_\Phi}{k} - 1 \right) = 2\pi \left( \sqrt{\dfrac{a}{a-6m}} -1 \right) \, ,
\end{align} 
where $a$ is the orbit's semi-major axis. Fig.\ \ref{fig_perigee} shows this precession of the perigee for different reference radii ranging from LEO to geostationary orbits. The result \eqref{eq_perigeePre1} is the same as shown in Ref.\ \cite{mTW:1973} and was also derived in \cite{Fuchs:1990} as well\footnote{Note the misprint in Eq.\ (4.14) in \cite{Fuchs:1990}, where actually the inverse value of the correct result is shown and we assume this to be simply a typo.}. Up to linear order in $m/a$ we obtain the well-known result
\begin{align}
	\Delta \varphi = \dfrac{6\pi m}{a} + \mathcal{O}\left( (m/a)^2 \right)  \, ,
\end{align}
that is the first term in Einstein's result \cite{Einstein:1916} 
\begin{align}
	\Delta \varphi = \dfrac{6\pi Gm}{a(1-e^2)} \approx \dfrac{6\pi m}{a} (1 + e^2 + e^4 + \dots ) \, ,
\end{align}
for the precession of the perigee in the case of small eccentricities. For the second kind of an elliptically perturbed orbit that is described by Eq.\ \eqref{eq_mixedOrbit} we obtain according to Eq.\ \eqref{eq_mixedOrbitPrecession}
\begin{multline}
	\Delta \varphi 	= \dfrac{2\pi (\Omega_\Phi + \delta \omega_{(1)})}{k} -2\pi \\= 2\pi \left( \sqrt{\dfrac{R}{R-6m}} -1 \right) 
					- 3\pi \dfrac{R-2m}{R-3m} \sqrt{\dfrac{R}{R-6m}} \dfrac{C_{(1)}}{R} \, .
\end{multline}
The first term resembles the previous result for the perigee precession where the parameter $C_{(1)}$ was set equal to zero and we recover this result in the limit. Up to linear order in $e$ and $m/a$ the result reads
\begin{align}
	\Delta \varphi = \dfrac{6\pi m}{a} \left( 1 - \left( 1+\dfrac{m}{2a} \right) e \right) +  \mathcal{O}\left(e^2, (m/a)^2 \right)\, ,
\end{align}
and depends on the eccentricity, whereas in the first case the result was independent of the perturbation parameters.

It should be mentioned that in the Newtonian solution \eqref{eq_NewtonDevSolOrbit} no perigee precession is present since there is only one frequency, the Keplerian frequency $\Omega_K$, involved. In Newtonian gravity (at least for a spherically symmetric potential) the Kepler ellipses are closed. Hence, as it is well-known, the precession of an elliptical orbit is a relativistic effect. It is recovered in the framework of the first order geodesic deviation equation due to the appearance of a second frequency in the relativistic solution.


\subsection{Redshift and Time Dilation}
The redshift $z$ between two standard clocks that show proper times $s$ and $\tilde{s}$ is
\begin{align}
	\label{Eq_1}
  	1 + z = \dfrac{\nu}{\tilde{\nu}} = \dfrac{d\tilde{s}}{ds} \, .
\end{align}
Using the solution of the first order deviation equation, we can derive a formula for the redshift between the clocks transported along the reference and deviating orbit as follows. Along the orbit $X^a(s)$, the constant of motion $E_X$ related to the energy is
\begin{align}
	\label{Eq_2}
  	E_X = \left(1-\dfrac{2m}{r(\tilde{s})} \right) \dfrac{dt}{d\tilde{s}} = \left(1-\dfrac{2m}{r(s)} \right) \dfrac{dt}{ds} \dfrac{ds}{d\tilde{s}} \, .
\end{align}
Hence, we obtain for the redshift using the solution $t(s)$ and $r(s)$ of the first order deviation equation
\begin{align}
	\label{Eq_3}
  	z + 1 = \dfrac{d\tilde{s}}{ds} = \left(1-\dfrac{2m}{r(s)} \right) \dfrac{\dot{T} + \dot{\eta}^t(s)}{E_X} = \left(1-\dfrac{2m}{R + g(s)} \right) \dfrac{\sqrt{\dfrac{R}{R-3m}} + \dfrac{\sqrt{m R}}{R-2m}\dot{f}(s)}{E_X} \, ,
\end{align}
where the functions $f(s)$ and $g(s)$ are given by Eqns.\ \eqref{eq_GR_CircRefDevSol2} and $E_X$ is fixed by the initial conditions of the deviating orbit, i.e.\ by the choice of parameters $C_{(i)}$. 
\begin{multline}
	z+1 = \dfrac{1}{E_X} \left( 1 - \dfrac{2m}{R+C_{(1)}+C_{(2)}\sin ks + C_{(3)} \cos ks} \right) \\
	\times \left( \sqrt{\dfrac{R}{R-3m}} + \lambda_{(1)} + \lambda_{(2)} \sin ks + \lambda_{(3)} \cos ks \right) \, ,
\end{multline}
where
\begin{align}
	\lambda_{(1)} 	= -\dfrac{3}{2} \dfrac{\sqrt{m}(R-2m)}{R(R-3m)^{3/2}} \, C_{(1)} \, , \quad 
	\lambda_{(2,3)} = - 2 \dfrac{\sqrt{m}R}{(R-2m) \sqrt{R-6m}} \, C_{(2,3)} \, .
\end{align}
This result yields a comparatively simple model for the redshift between the two satellites and is accurate as long as the orbital deviation is small.

For two circular orbits with radii $R$ and $R+C_{(1)}$, we recover the correct result to first order in $C_{(1)}/R$. The redshift becomes
\begin{align}
	z = \dfrac{C_{(1)}}{R} \dfrac{3m}{2(R-m)} \, .
\end{align}
Note however, that terms related to Doppler effects are not present here, since we do not consider signals (light rays) send from one orbit to the other. Hence, the formula for the redshift contains only the part related to time dilation effects.


\subsection{Accuracy of the First Order Deviation Approach}

Exact solutions of the geodesic equation in the Schwarzschild spacetime can be constructed using elliptic functions. To our knowledge, the first work on this is contained in Refs.\ \cite{Forsyth:1920, Morton:1921}. The authors used the Jacobi elliptic functions $\text{sn}, \text{cn}, \text{dn}$ to solve the equation of motion. A more recent study of exact orbital solutions in Schwarzschild spacetime (and generalizations for, e.g., Kerr-Newman-deSitter spacetime) can be found in Refs.\ \cite{Hackmann:2008a, Hackmann:2008b, Hackmann:2009}, where the Weierstrass elliptic function $\Weierstrass$ is used.

Note that especially in the solutions of the geodesic equation that involve the Jacobi elliptic function $\text{sn}, \text{cn}, \text{dn}$ the relation to the solutions \eqref{eq_GR_CircRefDevSol} of the first order deviation equation is obvious. Take two such exact solutions and construct the deviation between the two orbits as their difference. Then, choosing one of these orbits to be the circular reference geodesic, the linearization of the deviation corresponds to the solution of the first order deviation equation. In Eq.\ \eqref{eq_GR_CircRefDevSol} $\sin$ and $\cos$ terms appear and these are the linearizations of the Jacobi elliptic functions $\text{sn}$ and $\text{cn}$.


\subsubsection{Circular Orbits}

\begin{figure}[htb]
\centering
	\includegraphics[width=0.75\textwidth]{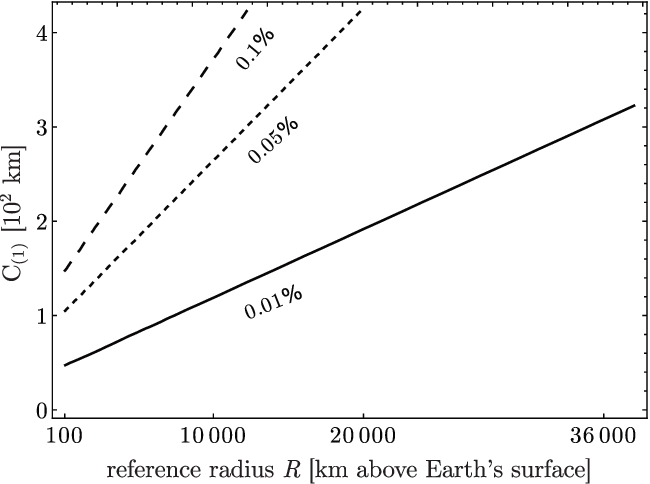}
	\caption{\label{fig_freqError} The dominating term $\delta \omega^{(2)}_{(1)}/\omega_\varphi$ in the frequency error. This relative error is made using the solution of the first order deviation equation to describe a circular perturbed orbit with initial radial distance $C_{(1)}$ to a reference orbit with radius $R$. Hence, the solution of the first order deviation equation gives the correct result up to a few parts in one hundred.}
\end{figure}

For a circular orbit in the equatorial plane with radius $r = R + C_{(1)}$, the exact azimuthal frequency is given by
\begin{align}
	\omega_\varphi = \sqrt{\dfrac{m}{(R+C_{(1)})^3}} \sqrt{\dfrac{R+C_{(1)}}{R+C_{(1)}-3m}} \, .
\end{align}
Expanding this result w.r.t.\ the small quantity ${C_{(1)}/R \ll 1}$ in a Taylor series, we find for the azimuthal frequency
\begin{align}
\label{eq_freqExpansion}
	\omega_\varphi 	&= \underbrace{ \sqrt{\dfrac{m}{R^3}} \sqrt{\dfrac{R}{R-3m}}}_{=\Omega_\Phi =: \delta \omega^{(0)}_{(1)} } \underbrace{ - \dfrac{3}{2} \dfrac{R-2m}{R-3m} \sqrt{\dfrac{m}{R^3}} \sqrt{\dfrac{R}{R-3m}} \dfrac{C_{(1)}}{R} }_{=:\delta \omega^{(1)}_{(1)}} \notag \\
	&\underbrace{ + \dfrac{15}{8} \dfrac{(R-2m)^2+4/5m^2}{(R-3m)^2} \sqrt{\dfrac{m}{R^3}} \sqrt{\dfrac{R}{R-3m}} \left( \dfrac{C_{(1)}}{R} \right)^2 }_{=:\delta \omega^{(2)}_{(1)}} \notag \\
	&+ \mathcal{O} \big( \left( C_{(1)}/R \right)^3 \big) \, .
\end{align}
The quantities $\delta \omega^{(i)}_{(1)}$ are defined as shown above. The superscript denotes the order of expansion and the subscript denotes the connection to the radial perturbation $C_{(1)}$. The $0^\text{th}$ order contribution is given by $\Omega_\Phi$, the azimuthal frequency for a circular orbit with radius $R$ - cf.\ Eq.\ \eqref{eq_TwoFrequencies}. Restricting ourselfes to first order contributions we can compare the approximation \eqref{eq_freqExpansion} to the solution of the first order deviation equation \eqref{eq_circPerturb}
\begin{align}
	\Omega_\Phi + \delta \omega_{(1)} \equiv \Omega_\Phi + \delta \omega^{(1)}_{(1)}.
\end{align}
Hence, the approximation up to linear order in $C_{(1)}/R$ is exactly the result that appeared in the solution of the first order deviation equation - cf.\ Eq.\ \eqref{eq_circPerturb}. Therefore, the error that we make using this result is dominated by the second order term $\delta \omega^{(2)}_{(1)}$. We show the relative error $\delta \omega^{(2)}_{(1)}/\omega_\varphi$ in Fig.\ \ref{fig_freqError} and conclude that even for a radial separation of some $10^2\,$km between the reference and perturbed orbit this error is less than $0.1\%$. Using the expansion in Eq.\ \eqref{eq_freqExpansion}, we estimate the error that remains at the next order, i.e.\ using also second order contributions in the frequency expansion. Then, the error is dominated by the third order term and about two orders of magnitude smaller. One can show that the quantity $\delta \omega^{(2)}_{(1)}/\omega_\varphi$, shown in Fig.\ \ref{fig_freqError}, is also the dominating error term for the conserved quantity $L$ that corresponds to the angular momentum of the perturbed orbit, since the frequency and angular momentum are related by $L = r^2 \omega_\varphi$.

Using the expansion \eqref{eq_freqExpansion}, we can write down the solution of the $n^\text{th}$ order deviation equation for a circular perturbation in the reference orbital plane. The perturbed orbit is then described by
\begin{subequations}
\label{eq_circPerturbOrderN}
\begin{align}
	r 		&= R+C_{(1)} \, ,\\
	\varphi &= \left( \Omega_\Phi + \delta \omega^{(1)}_{(1)} + \dots + \delta \omega^{(n)}_{(1)} \right) s \, ,
\end{align}
where
\begin{align}
	\delta \omega^{(k)}_{(1)} &= \dfrac{1}{k!} \frac{d^k \omega_\varphi}{d(C_{(1)}/R)^k} \left( \dfrac{C_{(1)}}{R} \right)^{k} \, .	
\end{align}
\end{subequations}
Table \ref{tab_CircPerturbError} shows the error that is made when modeling the distance between the reference orbit and the circular perturbed orbit after one complete reference period. To calculate the error we used the result above and compared it with the distance as calculated from two exact circular orbits with radii and frequencies $(R,\Omega_\Phi)$ and $(R+C_{(1)},\omega_\varphi)$.
\begin{table}[htb]
\centering
  \begin{tabular}{cccccc}
  \hline \hline
	 Radial separation & \multicolumn{4}{c}{Error [m]}   \\
	\hline
         $C_{(1)}$   & 1st & 2nd & 3rd & 4th  \\
                     & \multicolumn{4}{c}{order}  \\
    \hline 
    $R = R_\oplus + 1000\,$km & & & & \\
   $10\,$km & 159 & 0.25 & $\mathbf{4 \cdot 10^{-4}}$ & $\mathbf{6 \cdot 10^{-5}}$  \\
   $50\,$km & 3955 & 31 & 0.24 & \textbf{0.002}  \\
   $100\,$km & 15700 & 249 & 3.8 & \textbf{0.06}  \\
   $150\,$km & 35000 & 535 & 19 & 0.43  \\
   \hline
   $R = R_\oplus + 36000\,$km & & & & \\
   $10\,$km  & 27   & \textbf{0.007} 	& $\mathbf{10^{-5}}$ & $\mathbf{10^{-5}}$  \\
   $50\,$km  & 690  & 0.95 				& \textbf{0.001} & $\mathbf{5 \cdot 10^{-5}}$  \\
   $100\,$km & 2760 & 7.6 				& \textbf{0.02} & $\mathbf{5 \cdot 10^{-5}}$  \\
   $150\,$km & 6206 & 26 				& 0.1 & \textbf{0.0015}  \\
   \hline\hline
  \end{tabular}
  \caption{\label{tab_CircPerturbError} Error in the distance between reference and perturbed orbit after one complete reference period. We compare the distance as modeled with Eq.\ \eqref{eq_circPerturbOrderN}, up to $4^\text{th}$ order, with the distance constructed from two exact circular orbits with radii and frequencies $(R,\Omega_\Phi)$ and $(R+C_{(1)},\omega_\varphi)$. The table shows the values for the two different reference radii, $1000\,$km and $36000\,$km above Earth's surface. Bold marked values correspond to cm accuracy level when using the respective approximation order.}
\end{table}

\subsubsection{Elliptical Orbits}

To judge the accuracy of the elliptical orbits constructed with the solution \eqref{eq_elliptPerturb}, we compare them to exact solutions of the geodesic equation. In both cases we have to use identical initial conditions, i.e.\ the constants of motion $E$ and $L$ have to be the same. The exact orbit can either be constructed as a numerical solution of the geodesic equation, or by using analytic solutions in terms of elliptic functions.

\begin{figure}[htb]
\centering
	\includegraphics[width=0.75\textwidth]{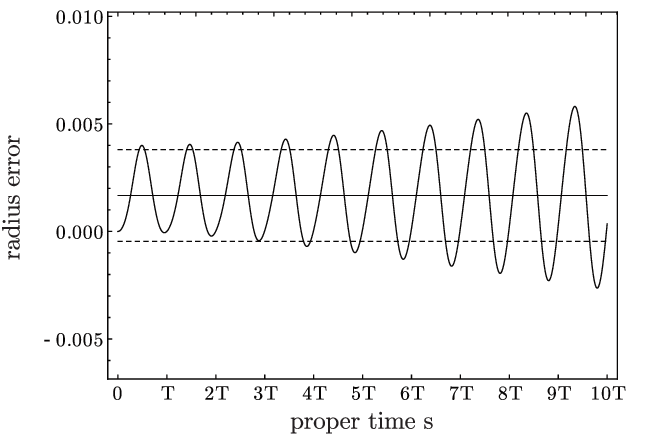}
	\includegraphics[width=0.75\textwidth]{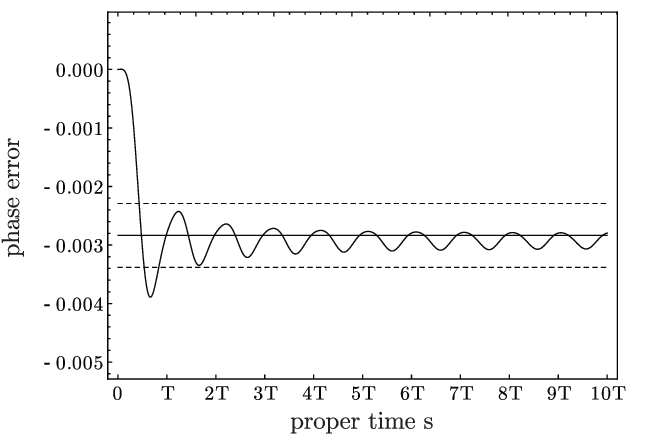}
	\caption{\label{fig_ellipticErrors} Relative error in the radial and azimuthal motion compared to the exact solution for 10 periods $(10T)$. The mean value (solid line) and range of one standard deviation around the mean (dashed lines) are shown.}
\end{figure}

We choose as initial conditions for eccentricity, distance to the perigee and argument of the perigee
\begin{align}
	e = 0.02 \, , \quad d_p = R_\oplus + 9672.6 \, \text{km}	 \, , \quad \omega = 0 \, .
\end{align}
The relative error in the radial and azimuthal motion is shown in Fig.\ \ref{fig_ellipticErrors} for ten orbital periods. The mean error in the radial deviation is positive, whilst the mean phase error is negative. Lowering the eccentricity by a factor of 10 yields relative errors that decrease two orders of magnitude. The analysis shows that the errors scale roughly as $e^2$, which is the expected behavior since terms that are quadratic in the eccentricity are neglected in this framework and contribute to second order deviations.

\subsubsection{Pendulum Orbits}
\begin{figure}[htb]
\centering
	\includegraphics[width=0.75\textwidth]{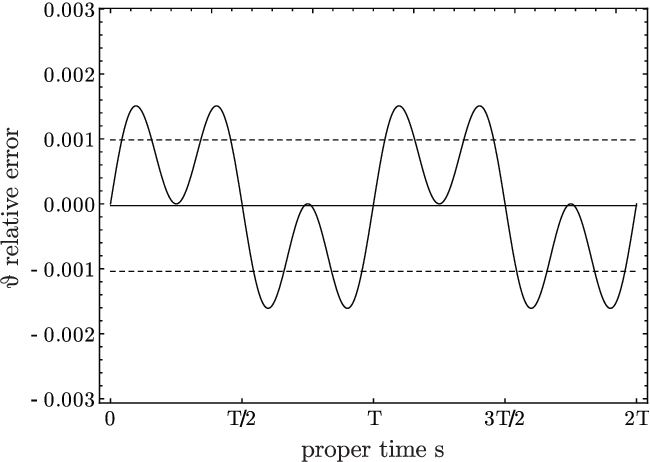}
	\caption{\label{fig_pendulum_thetaError} Relative error in the $\vartheta$-motion that is made using the solution \eqref{eq_inclinedPerturb} of the first order deviation equation, which describes a pendulum orbit constellation with an inclination of $5\, \deg$ between both orbital planes. The mean value (solid line) and one standard deviation around the mean (dashed lines) are shown as well.}
\end{figure}

Pendulum orbit constellations are of special interest for satellite geodesy missions. Hence, we should give an idea of how accurate the modeling of these constellations can be done using the solutions \eqref{eq_inclinedPerturb} of the first order geodesic deviation equation. Eq. \eqref{eq_inclinedPerturb} describes a circular orbit with unchanged radius that is inclined w.r.t\ the reference orbit. Thus, all constants of motion are the same but the orbital planes differ. In this solution, the amplitude $C_{(6)}/R$ of the $\vartheta$-oscillation gives the inclination of the perturbed orbit. Since we work with first order perturbations, this amplitude has to be small. For a reference radius of GRACE-type, $500\,$km above Earth's surface, and an inclination of $1\, \deg$ the error is about $100\,$ marcs. For two orbital periods, we show in Fig.\ \ref{fig_pendulum_thetaError} the relative error for an even higher inclination of $5\, \deg$. As can be seen in the figure, the error in the $\vartheta$-motion is periodic with zero mean value and standard deviation around $0.1 \, \%$ for this situation.

\subsection{Other Effects \label{sec_unphysEffects}}


\subsubsection{The Line of Nodes}

In Schwarzschild spacetime, due to symmetry, any orbit is confined to one orbital plane that is determined by the initial conditions. Conventionally this plane is then chosen to be the equatorial plane, defined by $\vartheta = \pi/2,\,  \dot{\vartheta} = \ddot{\vartheta} = 0$. Let there be two different bound orbits in Schwarzschild spacetime that we describe in the coordinate basis. We define the equatorial plane as the plane of motion for one of them. The second orbit is, of course, confined to a plane as well but this orbital plane is inclined w.r.t.\ the first one. Define the line of nodes as the spatial intersection of these two planes, i.e.\ the connection between the points where the second orbit crosses $\theta = \pi/2$. Since each of the two orbits is confined to its orbital plane, the line of nodes remains unchanged for an arbitrary number of revolutions.

We will now describe these two orbits using the solution \eqref{eq_GR_CircRefDevSol} of the first order deviation equation. The circular reference orbit with radius $R$ defines the equatorial plane. We choose $C_{(2,\dots,5)} \equiv 0$. Hence, only the parameters $C_{(1,6)}$ describe the shape of the perturbed orbit. As our analysis in the previous sections has shown, $C_{(1)}$ causes a constant radial perturbation and $C_{(6)}$ inclines the perturbed orbital plane. This perturbed orbit is described by
\begin{subequations}
\label{eq_LoNOrbit}
\begin{align}
	r 			 &= R + C_{(1)} \, , \\	
	\varphi(s) 	 &= ( \Omega_\Phi + \delta \omega_{(1)} ) \,s \, , \\
	\vartheta(s) &= \pi/2 + \dfrac{C_{(6)}}{R} \sin \Omega_\Phi s \, .
\end{align}
\end{subequations}
The line of nodes is determined by two successive intersections of the $\vartheta$-motion with the equatorial plane. This happens for $s = n \pi/\Omega_\Phi \,,~n=0,1,2..$. For these values we get
\begin{subequations}
\begin{align}
	\vartheta(s = n \pi/\Omega_\Phi) &= \pi/2, \\
	\varphi(s = n \pi/\Omega_\Phi) &= n \pi + \dfrac{n \pi \delta \omega_{(1)}}{\Omega_\Phi} \, .
\end{align}
\end{subequations}
Thus, from one orbit to the next the line of nodes shifts by an amount of
\begin{align}
	\Delta n := \dfrac{2 \pi \delta \omega_{(1)}}{\Omega_\Phi} = - 3\pi \dfrac{R-2m}{R-3m} \dfrac{C_{(1)}}{R} \, .
\end{align}
For $C_{(1)}/R = 1\%$ this corresponds to about $-5\, \deg$. Since a precession of the line of nodes must not happen in Schwarzschild spacetime, we have to carefully analyze this kind of ``effect''. As we have shown in the expansion of the exact azimuthal frequency \eqref{eq_freqExpansion} for a circular orbit with radius $R + C_{(1)}$, $\Omega_\Phi$ is its $0^\text{th}$ order and $\delta \omega_{(1)}$ its $1^\text{st}$ order approximation. The approximation order in the frequency of the $\vartheta$-motion in \eqref{eq_LoNOrbit} is one less than the order in the $\varphi$-motion, because the amplitude $C_{(6)}/R$ of the $\vartheta$-motion is already of first order. Hence, when using higher order deviation solutions up to $j$th order, the orbit will be given by
\begin{subequations}
\begin{align}
	r &= R + C_{(1)} \, ,\\	
	\varphi &= (\Omega_\Phi + \delta \omega^{(1)}_{(1)} + \dots + \delta \omega^{(j)}_{(1)} )\, s \, , \\
	\vartheta &= \pi/2 + C_{(6)} \sin (\Omega_\Phi + \delta \omega^{(1)}_{(1)} + \dots + \delta \omega^{(j-1)}_{(1)} )\, s \, ,
\end{align}
\end{subequations}
and subsequently the shift of the line of nodes after one orbit is
\begin{align}
	\Delta n = \dfrac{2\pi \delta \omega^{(j)}_{(1)}}{\Omega_\Phi + \delta \omega^{(1)}_{(1)} + \dots + \delta \omega^{(j-1)}_{(1)}} \, .
\end{align}
Since $\delta \omega^{(j)}_{(1)}$ is the $j$th term in the Taylor expansion \eqref{eq_freqExpansion} we notice that $\Delta n \to 0$ for $j \to \infty$. Hence, the shift of the line of nodes vanishes in the limit of infinite accuracy and is simply an artifact of the linearization/approximation.

\begin{figure}[htb]
	\centering
	\includegraphics[width=0.7\textwidth]{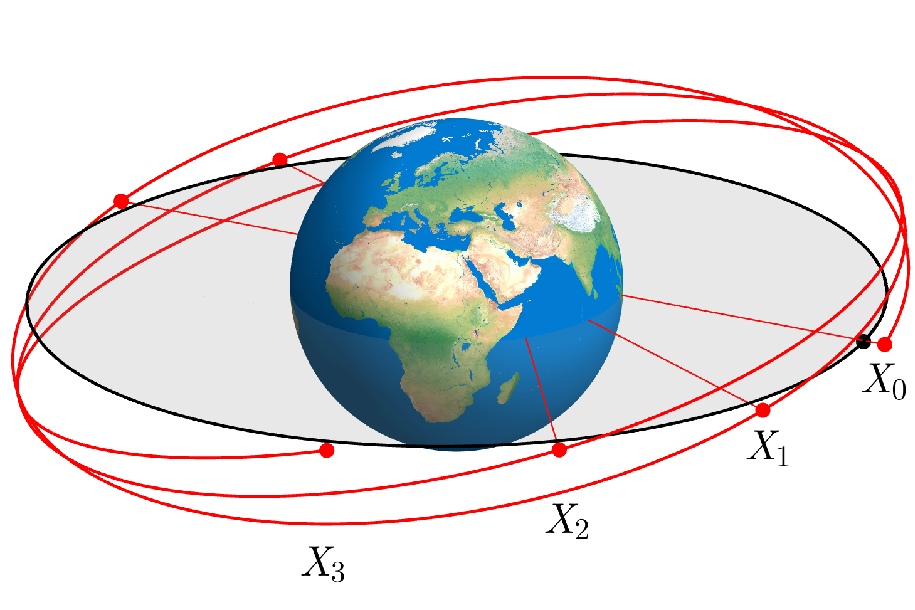}
	\caption{\label{fig_LoN} The reference and perturbed orbit for three reference periods. We have marked one endpoint of the line of nodes $X_n$ after $n$ elapsed reference periods. The precession of the perturbed orbital plane is clearly visible.}
\end{figure}


\subsubsection{Shirokov's Effect Revisited}

In \cite{Shirokov:1973} a new effect in the context of the standard geodesic deviation equation in Schwarzschild spacetime was reported. This effect was also studied in several follow-up works \cite{Nduka:1977, Vladimirov:etal:1981, Zimdahl:1985, Bergamin:etal:2009, Zimdahl:etal:2015}, in particular generalizations to other spacetimes than Schwarzschild were given and Shirokov's effect was compared to the influence of the oblateness of the Earth that causes similar perturbations of the reference geodesic.

In the following we critically reassess the derivation, meaning, and physical measurability of Shirokov's effect. Shirokov \cite{Shirokov:1973} did only consider periodic solutions of the standard geodesic deviation equation in Schwarzschild spacetime. The solution that is given in the work \cite{Shirokov:1973} is obtained in the framework presented here for $C_{(1,3,4,5)} = 0$. Only the parameters $C_{(2,6)}$ are involved. According to the previous analysis of the shape of the perturbed orbit this  corresponds to an elliptical orbit in an inclined orbital plane. This orbit is given by, cf.\ Eq.\ (21) in \cite{Shirokov:1973},
\begin{subequations}
\label{eq_ShirokovOrbit}
\begin{align}
	r(s) &= R + \eta^r(s) = R + C_{(2)} \sin ks \, , \\
	\varphi(s) &=\Omega_\Phi s + \eta^\varphi(s) = \Omega_\Phi s + \delta \omega_{(2)} \cos ks \, , \\
	\vartheta(s) &= \pi/2 + \eta^\vartheta(s) = \pi/2 + \dfrac{C_{(6)}}{R} \sin \Omega_\Phi s \, .
\end{align}
\end{subequations}
There are some subtleties that need to be handled with care. First of all, having identified a feature of a physical system within an approximate description, like the first order deviation equation, does not mean that this corresponds to a ``real'' effect. We have pointed out an example for this with the line of nodes precession that is present in the linearized framework but was shown to decrease with higher order approximations and to vanish in the end. Exactly this artifact of the line of nodes shift is present in Shirokov's solution as well. Since the equatorial plane is intersected by the perturbed orbit for $s=n \pi/ \Omega_\Phi$, we get a shift of the line of nodes after one $\vartheta$-period
\begin{align}
	\Delta \varphi = \varphi(s=2\pi/ \Omega_\Phi)	- 2\pi = \delta \omega_{(2)} \cos \dfrac{2\pi k}{\Omega_\Phi} \, .
\end{align}
This shift was not mentioned in the work by Shirokov, but is indeed present in the solution. However, Shirokov noticed correctly that in \eqref{eq_ShirokovOrbit} the $r$, $\varphi$ and $\theta$ oscillations have different frequencies and, thus, different periods
\begin{subequations}
\begin{align}
	T_r = T_\varphi &= 2\pi/k \, , \\
	T_\vartheta &= 2\pi/\Omega_\Phi \, .
\end{align}
\end{subequations}
Linearized in $m/R$ around the Keplerian value $T_K = 2\pi/\Omega_K = 2\pi \sqrt{R^3/m}$, we obtain
\begin{subequations}
\begin{align}
	T_r = T_\varphi &= T_K \left( 1 - \dfrac{3m}{R} \right) \, , \\
	T_\vartheta &= T_K \left( 1 + \dfrac{3m}{R} \right) \, .
\end{align}
\end{subequations}
It is by no means obvious why this linearization should be necessary, but it was used in \cite{Shirokov:1973}. Of course, for the Earth $m/R$ is a very small quantity when $R$ corresponds to radii above the surface, but the geodesic deviation equation works well even close to black holes, were $m/R$ might be large \cite{Koekoek:etal:2011}. However, Shirokov concludes that, due to the different periods of $\vartheta$ and $r$ oscillations, the distance $R\,(\vartheta-\pi/2)$ to the equatorial plane, in which the reference orbit lies, is different from $0$ after (several) radial oscillations and this is a new effect of GR. Shirokov imagines a satellite that moves on the reference orbit and rotates around the axis perpendicular to the equatorial plane with its orbital period, i.e.\ $2\pi/\Omega_\Phi$. Placed within this satellite a small test mass shall follow the perturbed orbit and the different frequencies of oscillations in the $\vartheta$ and $r$ direction can be observed.

Since each orbit in Schwarzschild spacetime is confined to its orbital plane, the $\vartheta$ and $\varphi$ frequencies have to be equal for exact orbits. Otherwise the line of nodes would shift. The difference in these periods that Shirokov discovered is a result of the approximation. However, the $r$ and $\varphi$ periods can be different, which leads to the perigee precession for elliptical orbits. Hence, the $r$ and $\vartheta$ periods can also be different for elliptical orbits to allow for a perigee precession within an inclined orbital plane. This is exactly what the approximate solution \eqref{eq_ShirokovOrbit} of the first order deviation equation describes: an elliptical orbit with perigee precession in an inclined orbital plane. The exact solution for this orbit would describe an elliptical orbit in an inclined but fixed orbital plane that shows perigee precession within this orbital plane. The solution \eqref{eq_ShirokovOrbit} is simply the first order approximation of this situation. Since the radial and the polar period are different, one would observe the object to be above or below the equatorial plane after (several) radial periods. This is nothing but a precessing ellipse in the inclined orbital plane. Hence, Shirokov's ``new'' effect is not new at all, it is the first order approximation of the well-known perigee precession -- discovered by Einstein already in 1916 \cite{Einstein:1916} -- in an inclined plane. Furthermore, as we have shown, this is mixed with the artifact of a precessing line of nodes due to the linearized framework.

We conclude that Shirokov's effect is no new effect but the first order description of the perigee shift for an elliptical orbit in an inclined orbital plane.


\section{Conclusions \label{sec_conclusion}}

We have shown how to describe orbits using the solution of the first order deviation equation for  circular reference geodesics. In particular, we employ the Schwarzschild spacetime as the simplest approximation for the Earth to investigate  relativistic satellite orbits and orbit deviations. We describe the shape of all perturbed orbits and connect free parameters in the general solution to the orbital elements of the perturbed orbit. Using this description, one can now apply the solution of the first order deviation equation to model any orbit that is specified in terms of its orbital elements. We have uncovered artificial effects that are due to the linearized framework. For elliptical orbits with small eccentricities the perigee precession was derived as a purely relativistic effect, which is absent in the Newtonian solution of the deviation problem. The solution of the first order deviation problem in Schwarzschild spacetime has shown that such an approximate description must be handled with care. The line of node precession, which is forbidden in Schwarzschild spacetime, will mix in the context of Kerr spacetime with the Lense-Thirring effect that causes a similar behavior. 

Reconsidering the so-called Shirokov effect we uncovered its origin. Rather than being a new feature of GR, we identified it as the relict of the approximate description of a well-known perigee precession.

The comparison of perturbed orbits -- based on the solution of the geodesic deviation equation -- to exact solutions of the underlying geodesic equation has shown, that higher order deviation equations should be used to model modern satellite based geodesy missions. 

In (simple) spacetimes, for which analytic solutions of the geodesic equation are available, one can estimate the accuracy of the approximate description to any order. In more realistic spacetimes numerical methods will become necessary.

In conclusion, already at the present level of accuracy applications in geodesy and gravimetry require the use of higher order deviation equations. Such equations will become indispensable for the  description of future high precision satellite and ground based measurements. 


\section*{Acknowledgments}
The present work was supported by the Deutsche Forschungsgemeinschaft (DFG) through the grant PU 461/1-1 (D.P.), the Sonderforschungsbereich (SFB) 1128 Relativistic Geodesy and Gravimetry with Quantum Sensors (geo-Q), and the Research Training Group 1620 Models of Gravity. We also acknowledge support by the German Space Agency DLR with funds provided by the Federal Ministry of Economics and Technology (BMWi) under grant number DLR 50WM1547.

The authors would like to thank V.\ Perlick, J.W.\ van Holten, and Y.N.\ Obukhov for valuable discussions.


\appendix


\section{Conventions \& Symbols \label{app_conventions}}

In the following we summarize our conventions, and collect some frequently used formulas. A directory of symbols used throughout the text can be found in Tab.\ \ref{tab_symbols}. 
The signature of the spacetime metric is assumed to be $(+,-,-,-)$. Latin indices $i,j,k,\dots$ are spacetime indices and take values $0 \dots 3$.
For an arbitrary $k$-tensor $T_{a_1 \dots a_k}$, the symmetrization and antisymmetrization are defined by
\begin{eqnarray}
T_{(a_1\dots a_k)} &:=& {\frac 1{k!}}\sum_{I=1}^{k!}T_{\pi_I\!\{a_1\dots a_k\}},\label{S}\\
T_{[a_1\dots a_k]} &:=& {\frac 1{k!}}\sum_{I=1}^{k!}(-1)^{|\pi_I|}T_{\pi_I\!\{a_1\dots a_k\}},\label{A}
\end{eqnarray}
where the sum is taken over all possible permutations of its $k$ indices, symbolically denoted by $\pi_I\!\{a_1\dots a_k\}$. 
 
\begin{table}[]
\centering
\begin{tabular}{ll}
\hline
\hline
Symbol & Explanation\\
\hline
&\\
\hline
\multicolumn{2}{l}{{Geometrical quantities}}\\
\hline
$g_{a b}$ & Metric\\
$A(r)$, $B(r)$ & Free metric functions \\
$\sqrt{-g}$ & Determinant of the metric \\
$\delta^a_b$ & Kronecker symbol \\
$x^{a}$, $s$ & Coordinates, proper time \\
$Y^{a}$, $X^{a}$ & (Reference, perturbed) curve \\
$\Gamma_{a b}{}^c$ & Connection \\
$R_{a b c}{}^d$& Curvature \\
$\eta^a$ & Deviation vector\\
\hline
\multicolumn{2}{l}{{Physical quantities}}\\
\hline
$G$, $U$ & Newtonian gravitational (constant, potential)\\
$R$ & Reference radius\\
$\Omega_K$, $T_K$ & Keplerian frequency, period\\
$\Omega_\Phi$, $\omega_\varphi$ & Azimuthal freq. of circ. orbit\\
$k$ & Second freq. in relativistic solution\\
$T_r$, $T_\varphi$, $T_\vartheta$ & Frequencies of perturbations\\
$C_{(i)}$ & Perturbation parameters, $i=1 \dots 6$\\
$\delta \omega_{(i)}$, $\delta t_{(i)}$ & (Azimuthal, temporal) deviations, $i=1\dots 3$\\
$R_\oplus$ & Mean Earth radius $6.37 \cdot 10^3\,$km\\
$\Delta \varphi$, $\Delta n$  & Angle of (perigee precession, line of nodes shift)\\
$E$ & Energy \\
$L$ & Angular momentum \\
$\rho$ & Matter density\\
$M$, $m$  & Mass of the central object [kg], [m]\\
$f(s)$, $g(s)$, $\Delta$ & Abbreviations \\
\hline
\multicolumn{2}{l}{{Orbital elements}}\\
\hline
$a$ & Semi major axis\\
$e$ & Eccentricity\\
$\Omega_a$ & Ascending node\\
$i$ & Inclination\\
$d_p$, $d_a$ & Distance to (perigee, apogee)\\
$\omega$ & Argument of the perigee\\
$\nu$ & True anomaly\\
\hline
\multicolumn{2}{l}{{Operators}}\\
\hline
$\partial_i~$, $\nabla_i$ & (Partial, covariant) derivative \\
$\frac{D}{ds} = $``$\dot{\phantom{a}}$'' & Total covariant derivative\\ 
\hline
\hline
\end{tabular}
\caption{\label{tab_symbols}Directory of symbols.}
\end{table}

The covariant derivative defined by the Riemannian connection is conventionally denoted by the nabla or by the semicolon: $\nabla_a =$ ``$ {}_{;a}$''. Our conventions for the Riemann curvature are as follows:
\begin{align}
&2 A^{c_1 \dots c_k}{}_{d_1 \dots d_l ; [ba] } \equiv 2 \nabla_{[a} \nabla_{b]} A^{c_1 \dots c_k}{}_{d_1 \dots d_l} \notag \\
&= \sum^{k}_{i=1} R_{abe}{}^{c_i} A^{c_1 \dots e \dots c_k}{}_{d_1 \dots d_l} - \sum^{l}_{j=1} R_{abd_j}{}^{e} A^{c_1 \dots c_k}{}_{d_1 \dots e \dots d_l}. \label{curvature_def}
\end{align}
\clearpage


\bibliographystyle{unsrt}


\end{document}